\documentclass[lettersize,journal]{IEEEtran}
\usepackage{amsmath,amsfonts}
\usepackage{algorithmic}
\usepackage{algorithm}
\usepackage{amsmath}
\usepackage{array}
\usepackage[caption=false,font=normalsize,labelfont=sf,textfont=sf]{subfig}
\usepackage{textcomp}
\usepackage{stfloats}
\usepackage{url}
\usepackage{verbatim}
\usepackage{graphicx}
\usepackage{cite}
\usepackage{amsmath,amssymb,amsfonts}
\usepackage{xcolor}
\usepackage{subcaption}
\usepackage{subfig} 
\usepackage{subfloat}
\usepackage{float}
\usepackage{stfloats}

\usepackage{siunitx}
\usepackage{acronym}
\usepackage {hyperref} 
\usepackage{ulem}

\acrodef{uav}[UAV]{unmanned aerial vehicle}
\acrodef{bs}[BS]{base station} 
\acrodef{nr}[NR]{new radio}
\acrodef{rcs}[RCS]{ radar cross section}
\acrodef{ofdm}[OFDM]{orthogonal frequency division multiplexing}
\acrodef{isac}[ISAC]{integrated
sensing and communication}
\acrodef{csi}[CSI]{channel state information} 
\acrodef{snr}[SNR]{signal-to-noise ratio}
\acrodef{mse}[MSE]{mean-squared error} 
\acrodef{stft}[STFT]{short-time Fourier transform}
\acrodef{ifft}[IFFT]{Inverse Fast Fourier Transform}
\acrodef{fft}[FFT]{Fast Fourier Transform}
\acrodef{pdcch}[PDCCH]{Physical Downlink Control Channel}
\acrodef{pdsch}[PDSCH]{Physical Downlink Shared Channel}
\acrodef{dl}[DL]{downlink}
\acrodef{ul}[UL]{uplink}
\acrodef{emd}[EMD]{empirical mode decomposition}
\acrodef{imf}[IMF]{intrinsic mode function}
\acrodef{vmd}[VMD]{variational mode decomposition}
\acrodef{nsp}[NSP]{Null space pursuit}
\acrodef{rmD-NSP}[rmD-NSP]{rotor micro-Doppler null space pursuit}
\acrodef{tdd}[TDD]{time division duplexing} 
\acrodef{nr}[NR]{New Radio} 
\acrodef{tdd}[TDD]{time division duplexing} 
\acrodef{pri}[PRI]{pulse repetition interval}
\acrodef{prf}[PRF]{pulse repetition frequency}
\acrodef{cpi}[CPI]{coherent processing interval}
\acrodef{set}[SET]{synchroextracting transform}
\acrodef{idft}[IDFT]{inverse discrete Fourier transform}
\acrodef{if}[IF]{instantaneous frequency}

\begin{document}

\title{UAV's Rotor Micro-Doppler Feature Extraction Using Integrated Sensing and Communication Signal: Algorithm Design and Testbed Evaluation}

\author{Jiachen~Wei, Dingyou~Ma, Feiyang~He,  Qixun~Zhang, Zhiyong~Feng, Zhengfeng~Liu, Taohong~Liang
  \thanks{This work is partly supported by Beijing Natural Science Foundation (L232003), National Natural Science Foundation of China (62321001, 62341101), Fundamental Research Funds for the Central Universities (No. 24820232023YQTD01) and Fundamental Research Funds for the Central Universities (2024RC02).}		
  \thanks{J. Wei, D. Ma, F. He, Q. Zhang, Z. Feng, Z. Liu, and T. Liang are with the Key Laboratory of Universal Wireless Communications, Ministry of Education, Beijing University of Posts and Telecommunications, Beijing 100876, China (Email:\{weijiachen, dingyouma\}@bupt.edu.cn, flyinghjgc@163.com, \{zhangqixun, fengzy\}@bupt.edu.cn, guan\_kangping@163.com, 13381221925@189.cn).}
  \thanks{(\textit{Corresponding authors: Dingyou~Ma})}
	}
\markboth{}%
{Shell \MakeLowercase{\textit{et al.}}: A Sample Article Using IEEEtran.cls for IEEE Journals}


\maketitle

\begin{abstract}
 With the rapid application of \acp{uav} in urban areas, the identification and tracking of hovering UAVs have become critical challenges, significantly impacting the safety of aircraft take-off and landing operations. As a promising technology for 6G mobile systems, \ac{isac} can be used to detect high-mobility UAVs with a low deployment cost. The micro-Doppler signals from UAV rotors can be leveraged to address the detection of low-mobility and hovering UAVs using ISAC signals. However, determining whether the frame structure of the ISAC system can be used to identify UAVs, and how to accurately capture the weak rotor micro-Doppler signals of UAVs in complex environments, remain two challenging problems.
 This paper first proposes a novel frame structure for UAV micro-Doppler extraction and the representation of UAV micro-Doppler signals within the channel state information (CSI). 
 Furthermore, to address complex environments and the interference caused by UAV body vibrations, the \ac{rmD-NSP} algorithm and the feature extraction algorithm \ac{set} are designed to effectively separate UAV's rotor micro-Doppler signals and enhance their features in the spectrogram. 
 Finally, both simulation and hardware testbed demonstrate that the proposed rmD-NSP algorithm enables the \ac{isac} base station (BS) to accurately and completely extract UAV's rotor micro-Doppler signals. Within a $\SI{0.1}{s}$ observation period, ISAC BS successfully captures eight rotations of the DJI M300 RTK UAV's rotor in urban environments. Compared to the existing AM-FM NSP and NSP signal decomposition algorithms, the integrity of the rotor micro-Doppler features is improved by 60\%.
\end{abstract}


\section{Introduction}
\IEEEPARstart{T}{h}e rapid development of 5G 
 mobile communication systems and artificial intelligence technology has drawn significant attention to the low-altitude economy~\cite{UAVbackground}. The low cost, small size, ease of acquisition, and particularly live video feed have greatly popularized the use of unmanned aerial vehicles (UAVs). Unfortunately, the increasing popularity of \ac{uav} applications has raised concerns about safety, security, and privacy, mainly due to the so-called smart attacks and threats~\cite{UAVdanger}.
 It is critically important to find ways to counter the illegal \acp{uav} intrusion~\cite{uavsecurity}. Therefore, numerous researchers from academia and industry have tried to develop suitable \ac{uav} detection systems, primarily through the use of multiple sensors, such as ultrasonic sensors, light detection and ranging (LiDAR), optical cameras, millimeter wave (mmWave) radars, etc. At the same time, 
 the next generation of wireless networks, in addition to providing primary communication functions, is also expected to gain the capability to sense the surrounding environment through radio frequency (RF) signals, effectively acting as a sensor~\cite{massiveMIMO}. Leveraging existing communication \acp{bs} and pipeline infrastructure, a \ac{uav} detection network can be rapidly and cost-effectively constructed. Furthermore, the inherent network collaborative capability of multiple \acp{bs} can enhance the detection accuracy and reduce blind spot detection, which can address the problem of \ac{uav} being difficult to detect in urban scenarios~\cite{CYP_UAV}. 
 
  
 As one of the key technologies for 6G communication, integrated sensing and communication (ISAC) systems provide theoretical feasibility for using mobile communication systems to detect UAVs~\cite{SensingNetwork}. Current research on \ac{isac} primarily focuses on four aspects: signal design~\cite{informationTheory,Tradeoff,weiAsurvey}, signal processing~\cite{Andrewzhangsignalprocessing,MaIMmodulation,AndrewBeamArrays}, protocol-based applications~\cite{802.11targetdetect,ZhangJSTSP,jikejia,ISACZQX}, and networking sensing~\cite{networkDetection11,networkDetection2,networkDetection3}. In scenarios where multiple \acp{bs} are used to detect \acp{uav}, current research mainly focuses on UAV localization and tracking~\cite{chenxu,ISACdetectiontarget,ISACdetectiontarget2}. Specifically, these \ac{isac} signal processing algorithms assume that the \ac{uav} is a point target with strong reflections and high-speed movement characteristics. These assumptions cannot align with the actual situations~\cite{UAVdifficultdetection,chen2019micro,GTD,reductionvibration}.
 First, the assumption that the UAV is a single scatterer neglects the reflections from the rotor blades. Second, UAV typically move at a slower speed and can hover, making them difficult to detect from clutter, noise, and other natural flying objects like birds in urban environments. 
 Actually, the rotation of the \ac{uav}'s rotor introduces unique high-frequency modulation sidebands near the Doppler frequency, known as the micro-Doppler signal. This micro-Doppler signal can be used to identify the type of target and distinguish it from other objects~\cite{Vc.chenbook}. 

 However, existing ISAC studies do not consider the application of micro-Doppler to detect and identify \ac{uav} in mobile communication systems. There is still a lack of practical performance evaluation for using mobile communication systems to extract micro-Doppler signatures of UAV. The challenges with micro-Doppler extraction of UAV based on mobile communication systems come from two main sources. One aspect is whether the communication waveform and uplink/downlink frame structure affect the micro-Doppler feature extraction of UAVs. On the other hand, 
 considering the complex environment faced by micro-Doppler signal extraction, there is a significant amount of clutter and interference. The UAV's rotor echoes to be extracted are very weak, making the extraction of micro-Doppler features challenging. Furthermore, the vibration of UAVs itself also affects the performance of micro-Doppler extraction~\cite{costa2024uavmodelling,UAVvibration1}.

 Therefore, to address the above issues, this paper presents a \ac{tdd} frame structure configuration for extracting micro-Doppler features of \ac{uav} using ISAC signal. Furthermore, it provides an expression for the echoes of UAV in the context of \ac{csi}. \textcolor{black}{This expression takes into account the time-varying \ac{rcs} caused by the UAV's rotor rotation, leading to periodic variations in rotor echo intensity.} Additionally, it considers the translational and vibrational motion models of the UAV body. 
 To combat the complex environment and suppress the effect of body vibrations, the rotor micro-Doppler null space pursuit (rmD-NSP) algorithm is designed, based on an operator-based signal decomposition method~\cite{pengNSP,PengNSP1,AMFMNSP}, to extract weak rotor micro-Doppler signals from various interference signals. 
 The rotor micro-Doppler signals obtained from decomposition can be further processed using the synchroextracting transform (SET) algorithm to obtain time-frequency ridges with more concentrated energy, thus revealing more pronounced micro-Doppler features~\cite{SET}. Finally, both simulation and hardware testbed results demonstrate that the proposed rmD-NSP algorithm can extract more UAV rotor micro-Doppler signals in urban environments compared to existing AM-FM NSP~\cite{AMFMNSP} and NSP~\cite{pengNSP} signal decomposition algorithms. The main contributions of this paper are summarized as follows.
 \begin{itemize}
\item  A new mathematical model for micro-Doppler of UAV in monostatic ISAC systems is proposed. This paper provides a model for the representation of the micro-Doppler signal in the \ac{csi} and considers the rotation of multiple rotor blades of UAVs, as well as the translational motion and vibration of the UAV body during flight. Additionally, the mathematical model takes into account the dynamic \ac{rcs} of rotors caused by the rotation of the blade, which will be an interesting tool for the future development of ISAC target classification algorithms.
\item We design a rmD-NSP adaptive operator based on the mathematical model of the UAV rotor. Through the zero space tracking algorithm, we extract rotor micro-Doppler signals from mixed signal containing many interference components. In addition, we use the time-frequency enhancement algorithm \ac{set} to complete the extraction of the micro-Doppler feature of UAVs. 
\item By conducting software simulations and hardware testbed based on the ISAC system, we validated the effectiveness of using the \ac{rmD-NSP} algorithm and \ac{set} algorithm to extract the micro-Doppler feature of UAVs in mobile communication systems. Within a $\SI{0.1}{s}$ observation period, ISAC BS successfully captures eight rotations of the DJI M300 RTK UAV's rotor in urban environment.
\end{itemize}

The rest of this paper is organized as follows. \textcolor{black}{Section~II presents the system model of the \ac{uav} micro-Doppler feature extraction.} Section~III introduces the rmD-NSP algorithm and the SET algorithm. In Section IV, we present the software simulation results. In Section V, we implement the extraction of micro-Doppler features from UAVs using the ISAC hardware testbed. \textcolor{black}{Section VI concludes this paper.}

The following notations are used throughout this paper: Boldface lowercase and uppercase letters denote vectors and matrices, respectively. We denote the transpose operation as $\left(\cdot\right)^{\text{T}}$. $\mathbf{A}_{\mathbf{x}} = \text{diag}(\mathbf{x})$ denotes the diagonalization of the vector $\mathbf{x}$. The set of complex numbers is $\mathbb{C}$ and the set of real numbers is $\mathbb{R}$.


\section{\textcolor{black}{SYSTEM MODEL}}

\begin{figure}[h]
    \centering
    
    \subfloat[]{\includegraphics[width=0.83\linewidth]{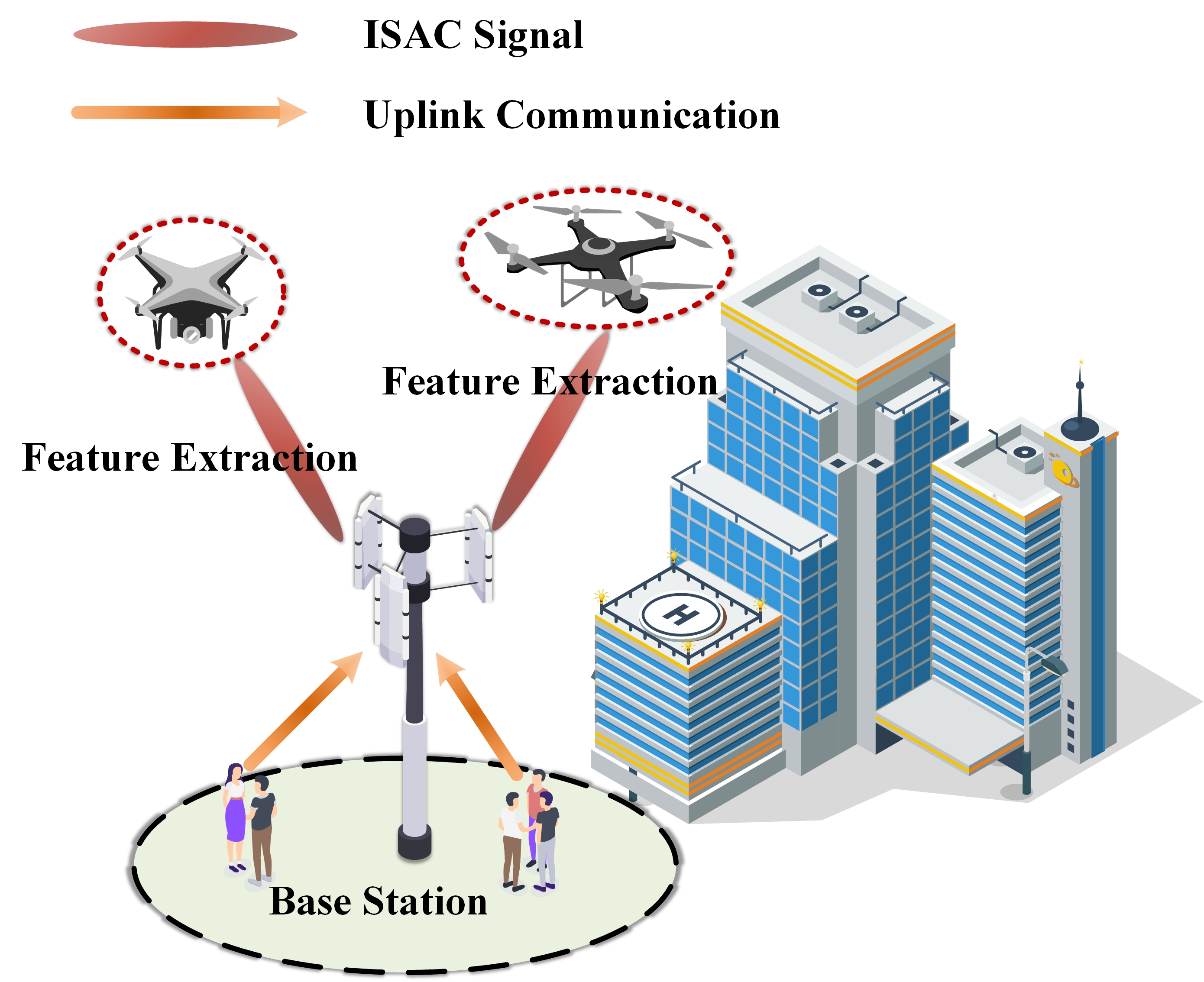}%
    }
    \hfil
    \subfloat[]{\includegraphics[width=0.83\linewidth]{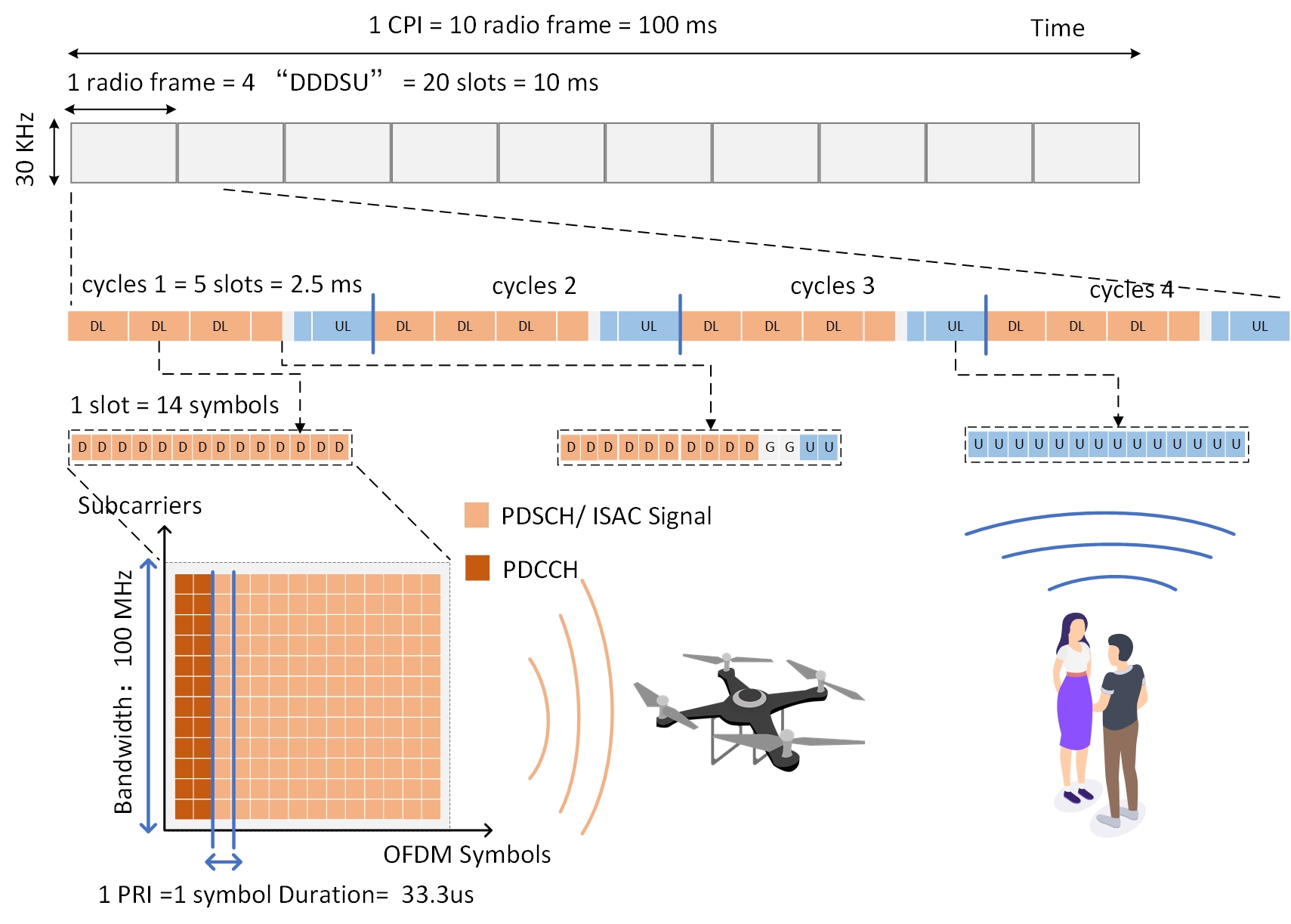}%
    }
    \caption{Low altitude UAV identification scenario and frame structure configuration: (a) Urban low altitude UAV identification scenario. (b) TD-ISAC Configuration.}
    \label{Fig:1}
\end{figure}

In this section, we consider an \ac{isac} \ac{bs} to implement the micro-Doppler feature extraction function for \ac{uav}, as shown in Fig.~\ref{Fig:1} (a). The \ac{bs} is equipped with separate antennas that enable full-duplex functionality, enabling it to transmit and receive signals simultaneously. 

First, in Subsection~II-A, we present a frame structure based on 3GPP TS 38.211~\cite{3GPP} for extracting UAV micro-Doppler features using \ac{bs}, as shown in Fig.~\ref{Fig:1} (b). In Subsection~II-B, we present the ISAC echo representation of the UAV, considering the dynamic \ac{rcs} characteristics and rotational motion of rotors, as well as the translational and vibrational motion of the UAV body. Then, in Subsection~II-C the \ac{csi} is given, which contains target range, Doppler and micro-Doppler information.
\subsection{5G NR TDD Configuration}

 The ISAC system proposed in this paper is based on the 5G \ac{nr} protocol. The 5G \ac{nr} 
 basic resource unit in the time domain is the radio frame, with each radio frame lasting for $\SI{10}{ms}$ and consisting of 10 subframes, each with a duration of $\SI{1}{ms}$. With the subcarrier spacing of $\SI{30}{kHz}$, each subframe contains 2 slots and each slot contains 14 \ac{ofdm} symbols. In the \ac{tdd} mode, there are three types of slots between \ac{bs} and UE: \ac{dl} slots, flexible slots and \ac{ul} slots. \ac{dl} slots are used for broadcasting by the \ac{bs} and transmitting control information and communication data to UE. The flexible slots serve as intervals for switching between \ac{ul} and \ac{dl} transmissions. \ac{ul} slots are used for UE feedback and information reporting to \ac{bs}. 

 Micro-Doppler feature extraction requires the sensing symbols in the ISAC system to be sufficiently dense. Therefore, we consider a frame structure designed for \ac{uav} micro-Doppler feature extraction as shown in Fig.~\ref{Fig:1} (b). During \ac{dl} slots, the \ac{bs} utilizes all \ac{pdsch} signals as sensing symbols, resulting in the shortest \ac{pri} of $\SI{33.3}{\mu\mathrm{s}}$. Currently, \ac{bs} operates in a \ac{tdd} mode with a single cycle duration of $\SI{2.5}{ms}$, i.e. ``DDDSU'', as the ISAC mode. After \ac{bs} detects an unknown flying target, it transmits communication signals to the target in the \ac{dl} slots while simultaneously receiving echoes. In this sensing mode, the extracted micro-Doppler features are unambiguous.

\subsection{\textcolor{black}{Received Signal Model}}
In sensing applications, we select several radio frames as one \ac{cpi}. During a single \ac{cpi}, the \ac{bs} transmits a total of $M$ \ac{pdsch} \ac{ofdm} symbols during \ac{dl} slots. The transmitted baseband signal $s\left(t\right)$ within a single CPI is expressed as
\begin{equation}
\begin{aligned}
s\left(t\right) =& \sum_{m=1}^{M}\sum_{n=1}^{N}\textcolor{black}{\mathbf{D_{\mathrm{TX}}}\left (  n,m\right )} 
\\
\textcolor{black}{\times} &\exp \left ( j 2\pi n\Delta f t \right )\mathrm{rect}\left(\frac{t-mT_{s}}{T_{s}}\right ),
\end{aligned}
\end{equation}
where $N$ denotes the number of subcarriers. \textcolor{black}{$\mathbf{D}_{\mathrm{TX}}  \in \mathbb{C}^{N \times M}$} represents the resource grid that transmits signals and \textcolor{black}{$\mathbf{D}_{\mathrm{TX}}\left(n,m\right)$} is the $m$ symbol modulated on the $n$ subcarrier, also called the complex modulation symbol. $\Delta f$ is the subcarrier interval and $T_{s} =\frac{1}{\Delta f}+T_{g}$ denotes the total duration of one OFDM symbol including that of the cyclic prefix (CP) $T_{g}$. $\mathrm{rect}\left(\cdot\right)$ is a rectangular window of unity support.

The \ac{isac} transmission signal propagates in free space and is reflected by the target. The transmitted signal in the time domain is $s\left(t\right)\exp\left(j2\pi f_{c}t\right)$ and $f_{c}$ represents the central frequency of the transmitter. Considering that there is a strong scattering point in the echos, the received signal $\widetilde{y} \left(t,R\left(t\right),\gamma\left(t\right)\right)$ at the receiving end is
\begin{equation}
\begin{aligned}
\widetilde{y} \left(t,R\left(t\right),\gamma\left(t\right)\right)=&\gamma\left(t\right) s\left(t-\frac{2R\left(t\right)}{c}\right)\exp\left(j2\pi f_{c}t\right)
\\
\textcolor{black}{\times} & \exp\left(-j2\pi f_{c} \frac{2R\left(t\right)}{c}\right),
\end{aligned}
\end{equation} 
where $\gamma\left(t\right)$ is the channel attenuation and $R\left(t\right)$ is the distance from the scattering point to the \ac{bs}.

\textcolor{black}{Scatterers on the UAV have different motion characteristics and scattering intensity characteristics. The scatterers from the rotor exhibit both translational and rotational motions. The rotational motion causes dynamic variations in the \ac{rcs}, leading to time-varying scattering intensities~\cite{Vc.chenbook}. The scatterers from the UAV body exhibit both translational and vibrational motions~\cite{UAVvibration1}}. The analysis is as follows.


The geometry of the ISAC BS and the UAV rotor is shown in Fig.~\ref{Fig:2}. For simplicity, the UAV is composed by a set of point scatterers, which are the primary reflecting points on the target. The target depicted in Fig.~\ref{Fig:2} represents one of the UAV blades. The BS is located at the origin $\left(X, Y, Z \right)$ of the space-fixed coordinates. The rotor is described in a local coordinate system $\left(x, y, z \right)$ that is attached to the target and has translation and rotation with respect to the BS coordinates. To observe the rotation of the rotor, a reference coordinates $\left(X^{\prime}, Y^{\prime}, Z^{\prime} \right)$ is introduced,  which shares the same origin with the target local coordinates and thus has the same translation as the target but no rotation with respect to the BS coordinates. The distance from the \ac{bs} to the \ac{uav} origin is $R_{0}$. The azimuth and elevation angles of the reference coordinate origin observed by the BS are $\alpha$ and $\theta$, respectively.

The rotor has a translation velocity $v$ relative to the BS and rotates around the Z-axis with a rotational speed of $f_r$. 
Thus, a point scatter $p$ in the UAV rotor at time $t = 0$ will move to $p^{\prime \prime}$ at time $t$. The movement consists of two steps: (1) translation from $p$ to $p^{\prime}$, as shown in Fig.~\ref{Fig:2}, with a velocity $v$; and (2) rotation from $p^{\prime}$ to $p^{\prime \prime}$ with a rotational speed of $f_r$. The modeling of the motion characteristics and scattering properties of different scatterers on the \ac{uav} is as follows:
\begin{figure}[h]
    \centering
    \includegraphics[width=0.95\linewidth]{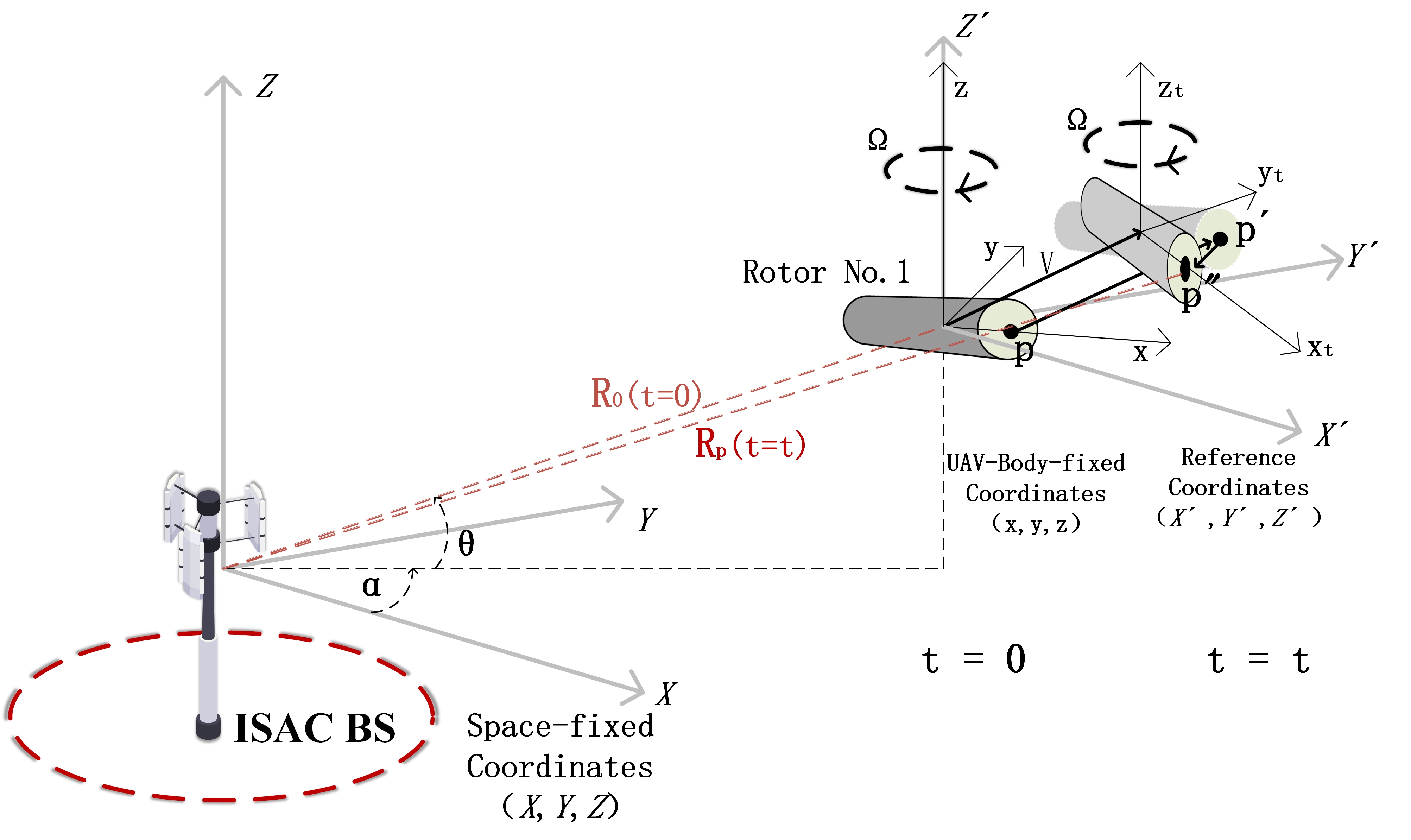}
    \caption{Gemotry of the BS and the UAV first rotor with translation and rotation.}
    \label{Fig:2}
\end{figure}
\subsubsection{UAV Body Translation}
The body of the UAV can be regarded as a strong scattering point. The radial motion pattern of the scatterer on the body, representing translational movement relative to the \ac{bs}, is as follows:
\begin{equation}
    R_{\text{trans}}^{\text{body}}\left(t\right) = R_{0} + vt.
\end{equation}
During a single \ac{cpi}, the radial motion distance of the UAV is very short. Hence, its scattering intensity $\gamma^{\text{body}}\left(t\right)$ can be considered as a constant $\gamma^{\text{body}}$. Therefore, the echo from the scatterer due to the translational motion of the body can be expressed as 
\begin{equation}
y_{\text{trans}}^{\text{body}}\left(t\right) = \widetilde{y} \left(t,R_{\text{trans}}^{\text{body}}\left(t\right),\gamma^{\text{body}}\right).
\end{equation}
\subsubsection{UAV Rotor Rotation}
Each UAV is equipped with multiple blades, one of which is illustrated in Fig.~\ref{Fig:2}.
Ignoring the distance from the UAV blade tip to the UAV body. Let azimuth angle $\alpha = 0$, the distance from the BS to the scatter point $p$ (at the tip of the blade) can be expressed as
\begin{equation}
\begin{aligned}R_{p}^{\text{rotor}}\left(t\right) = R_{0} + vt + \frac{L}{2} \cos \theta \cos\left(2 \pi f_r t + \varphi_{p}\right), 
\end{aligned}
\end{equation}
where $L$ is the length of the blade, and $\left(L / R_0\right)^2 \rightarrow 0$ in the far field. $\varphi_{p}$ represents the initial rotation angle of the scattering point $p$ on the blade. 

The high-speed rotation of the blades results in a dynamic \ac{rcs}~\cite{rcs}. To simplify matters, we can allocate the \ac{rcs} of each blade to the blade's tip~\cite{Vc.chenbook}. The \ac{rcs} of scatterer $p$ at time $t$ can be expressed as
\begin{equation}
\begin{aligned}
\sigma_{p}^{\text{rotor}}\left(t\right) &= \sum_{\ell =1}^{\infty} \sum_{i=1}^{I} a_i \sin \left[ b_i \frac{f_r}{100} \left( t + \frac{\varphi_{p}}{2\pi f_r} \right) + c_i \right]\\
&\times u_1 \left( t - \frac{\ell }{f_r} + \frac{\varphi_{p}}{2\pi f_r} \right),
\end{aligned}
\end{equation}
where
\begin{equation}
    u_{1}\left(t\right)=\left\{\begin{array}{ll}1, & 0 \leq t<\frac{1}{f_r} \\0, & t>\frac{1}{f_r}\end{array}\right.,
\end{equation}
and $I$ varies according to the material of the UAV blades. $a_i,b_i$ and $c_i$ represent the coefficients under different materials, respectively. Within a single \ac{cpi}, this dynamic \ac{rcs} results in rapid variations in the scattering intensity $\gamma_{p}^{\text{rotor}}\left(t\right)$. Considering multiple scatterers on multiple rotors, the echo from the scatterers due to the rotor rotation can be expressed as 
\begin{equation}
y^{\text{rotor}}_{\text{rotation}}\left(t\right) = \sum_{p=1}^{P}\widetilde{y} \left(t,R_{p}^{\text{rotor}}\left(t\right),\gamma_{p}^{\text{rotor}}\left(t\right)\right),
\end{equation}
where $P$ represent the total number of blades.

\subsubsection{UAV Body Vibration}
In practical applications, due to the interaction between the UAV body and the air, the UAV body has some high-frequency vibration components, which are added to the echoes from the target~\cite{vibration1}. The vibration components also generate micro-Doppler signals, which adversely affects the extraction of rotor micro-Doppler signal features.
\begin{figure}[h]
    \centering
    \includegraphics[width=0.95\linewidth]{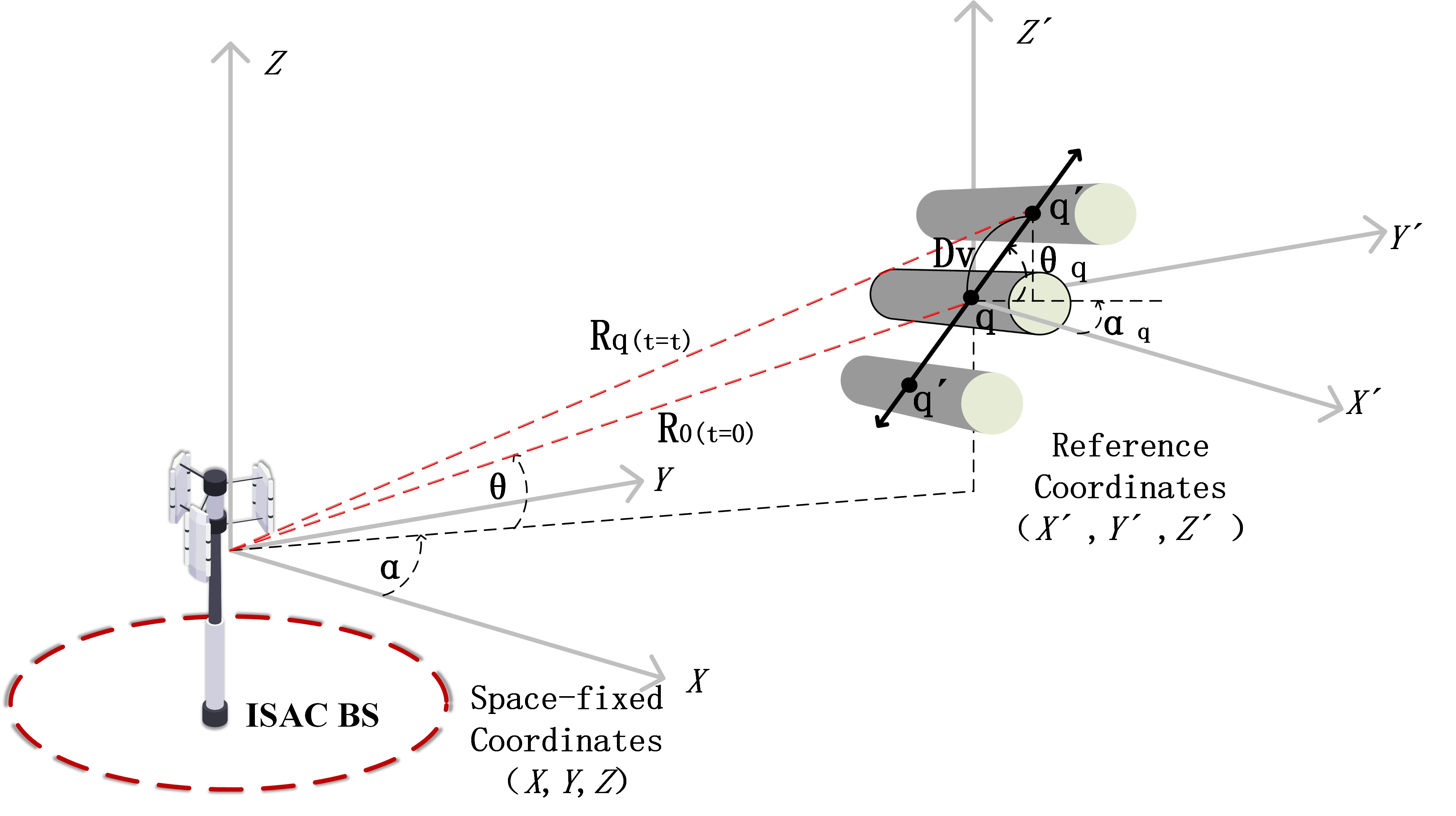}
    \caption{Gemotry of the BS and the UAV vibration. }
    \label{Fig:3}
\end{figure}

Assume that the scatter $q$ in the body of the UAV vibrates at a frequency $f_{v}$ with an amplitude $D_{v}$ and that the azimuth and elevation angle of the vibration direction in the reference coordinates $\left(X^{\prime}, Y^{\prime}, Z^{\prime} \right)$ are $\alpha_{q}$ and $\theta_{q}$, respectively, as shown in Fig.~\ref{Fig:3}. If the azimuth angle $\alpha$ and the elevation angle $\theta_{q}$ of the scatter $q$ are all zero, then we have
\begin{equation}
\begin{aligned}
R_q^{\text{body}}\left(t\right)=  R_0 + vt + D_v\sin\left(2\pi f_{v}t\right) \cos \theta \cos \alpha_q.
\end{aligned}
\end{equation}

The vibration of the \ac{uav} body has a minimal impact on the \ac{rcs}~\cite{reductionvibration}. Therefore, the scattering intensity of this scatterer $\gamma_{q}^{\text{body}}\left(t\right)$ can also be considered as a constant $\gamma^{\text{body}}$. The echo from the scatterer due to the vibration motion of the body can be expressed as 
\begin{equation}
y_{\text{vibration}}^{\text{body}}\left(t\right) = \widetilde{y} \left(t,R_q^{\text{body}}\left(t\right),\gamma^{\text{body}}\right).
\end{equation}

The total scattering echo of the \ac{uav} is given by the coherent sum of all independent scattering centers at that instant,
\begin{equation}
\begin{aligned}
\textcolor{black}{y\left(t\right) = y_{\text{trans}}^{\text{body}}\left(t\right) + y_{\text{vibration}}^{\text{body}}\left(t\right) + y^{\text{rotor}}_{\text{rotation}}\left(t\right) + n\left(t\right),}
\end{aligned}
\end{equation}
where $n\left(t\right)$ is additive white Gaussian noise (AWGN) with variance of $\sigma_{n}^2$.

\subsection{UAV CSI Model}


\textcolor{black}{The signal in (11) is sampled at the receiver with the period of $T_{s}$. Therefore, the sampled echo can be expressed as $y\left(t_{m}\right)$, where $t_{m} = mT_{s} \in  \left \{T_{s},2T_{s},\cdot\cdot\cdot,MT_{s}\right \}$ represents the discrete time. After RF (radio frequency) demodulation and OFDM demodulation, according to (1) and (11), the signal expression on the resource grid at the receiver is,}
\begin{equation}
   \mathbf{D}_{\mathrm{RX}} \left(n,m\right) = \mathbf{D}_{\mathrm{TX}}\left(n,m\right)\mathbf{\Theta } \left(n,m\right)+\mathbf{\Omega} \left(n,m\right),
\end{equation}
where $\mathbf{D}_{\mathrm{RX}} \in \mathbb{C}^{N \times M}$. $\mathbf{\Omega} \left(n,m\right)$ represents Gaussian white noise. The matrix $\mathbf{\Theta } = \boldsymbol{k}_{R} \otimes \boldsymbol{k}_{D}$, where $\boldsymbol{k}_{R}\in \mathbb{C}^{N\times1}$ carries the range information of the target, and  $\boldsymbol{k}_{D}\in \mathbb{C}^{1\times M}$ carries the Doppler and micro-Doppler information of the target. $\boldsymbol{k}_{R} \left(n\right)$ and $ \boldsymbol{k}_{D} \left(m\right)$ are modeled as follows:
\begin{equation}
    \boldsymbol{k}_{R} \left(n\right) = \exp\left(-j 2\pi  n\Delta f \frac{2R_{0}}{c}\right), n = 1,2,\cdots,N,
\end{equation}
\begin{equation}
\begin{aligned}
    \boldsymbol{k}_{D} \left(m\right) = \boldsymbol{k}_{D}^{\text{trans}} \left(m\right) + \boldsymbol{k}_{mD}^{\text{vibration}}\left(m\right) + &\boldsymbol{k}_{mD}^{\text{rotation}}\left(m\right), \\
    &m = 1,2,\cdots,M,
\end{aligned}
\end{equation}
where $\boldsymbol{k}_{D}^{\text{trans}}\in \mathbb{C}^{1 \times M}$ represents the Doppler phase information
due to the translational motion of the UAV body. $\boldsymbol{k}_{mD}^{\text{vibration}}\in \mathbb{C}^{1 \times M}$ represents the micro-Doppler phase information due to the body vibration, and $\boldsymbol{k}_{mD}^{\text{rotation}}\in \mathbb{C}^{1 \times M}$ represents the micro-Doppler amplitude-phase information due to the rotor rotation. The representations of $\boldsymbol{k}_{D}^{\text{trans}}$, $\boldsymbol{k}_{mD}^{\text{vibration}}$ and $\boldsymbol{k}_{mD}^{\text{rotation}}$ are as follows:
\begin{equation}
\begin{aligned}
&\boldsymbol{k}_{D}^{\text{trans}} \left(m\right) = \gamma^{body} \exp\left(-j2\pi f_{D}\left(m\right)\right),\\    
&\boldsymbol{k}_{mD}^{\text{vibration}}\left(m\right) = \gamma^{body}\exp\left(-j2\pi \left(f_{D}\left(m\right)+f_{mD}^{v}(m)\right)\right),\\
&\boldsymbol{k}_{mD}^{\text{rotation}}\left(m\right)= \sum_{p=1}^{P}\gamma^{rotor}_{p}\left(m\right)\\
&\quad\quad\quad\quad\quad\quad\quad\;\;\exp\left(-j2\pi \left(f_{D}\left(m\right)+f_{mD}^{p}\left(m\right)\right)\right),\\
&\quad\quad\quad\quad\quad\quad\quad\;\;\quad\quad\quad\quad\quad\quad\quad\;\;m = 1,2,\cdots,M,
\end{aligned}
\end{equation}
where $f_{D}\left(m\right) = \frac{2vmT_{s} }{\lambda}$ represents the Doppler effect caused by the translational motion. $f_{mD}^{v}\left(m\right) = \frac{2D_v \cos \theta \cos \alpha_q \sin\left(2\pi f_{v}mT_{s}\right)}{\lambda}$ represents the micro-Doppler effect caused by the vibration, and $f_{mD}^{p}\left(m\right) = \frac{ L \cos \theta \cos\left(2\pi f_r mT_{s} + \varphi_{p}\right)}{\lambda}$ represents the micro-Doppler effect caused by the rotor rotation.

Divide the corresponding elements of the transmitted signal on the resource grid to eliminate the influence of the amplitude and phase of the data. The signal model of the $\mathbf{CSI}$ containing target motion information is established as follows:
\begin{equation}
    \mathbf{CSI}\left(n,m\right) = \frac{\mathbf{D}_{\mathrm{RX}}\left(n,m\right)}{\mathbf{D}_{\mathrm{TX}}\left(n,m\right)} = \mathbf{\Theta }\left(n,m\right)+\mathbf{\Omega}^{\prime} \left(n,m\right).
\end{equation}

\section{UAV Rotor Feature Extraction Algorithm}

\begin{figure}[h]
    \centering    \includegraphics[width=0.9\linewidth]{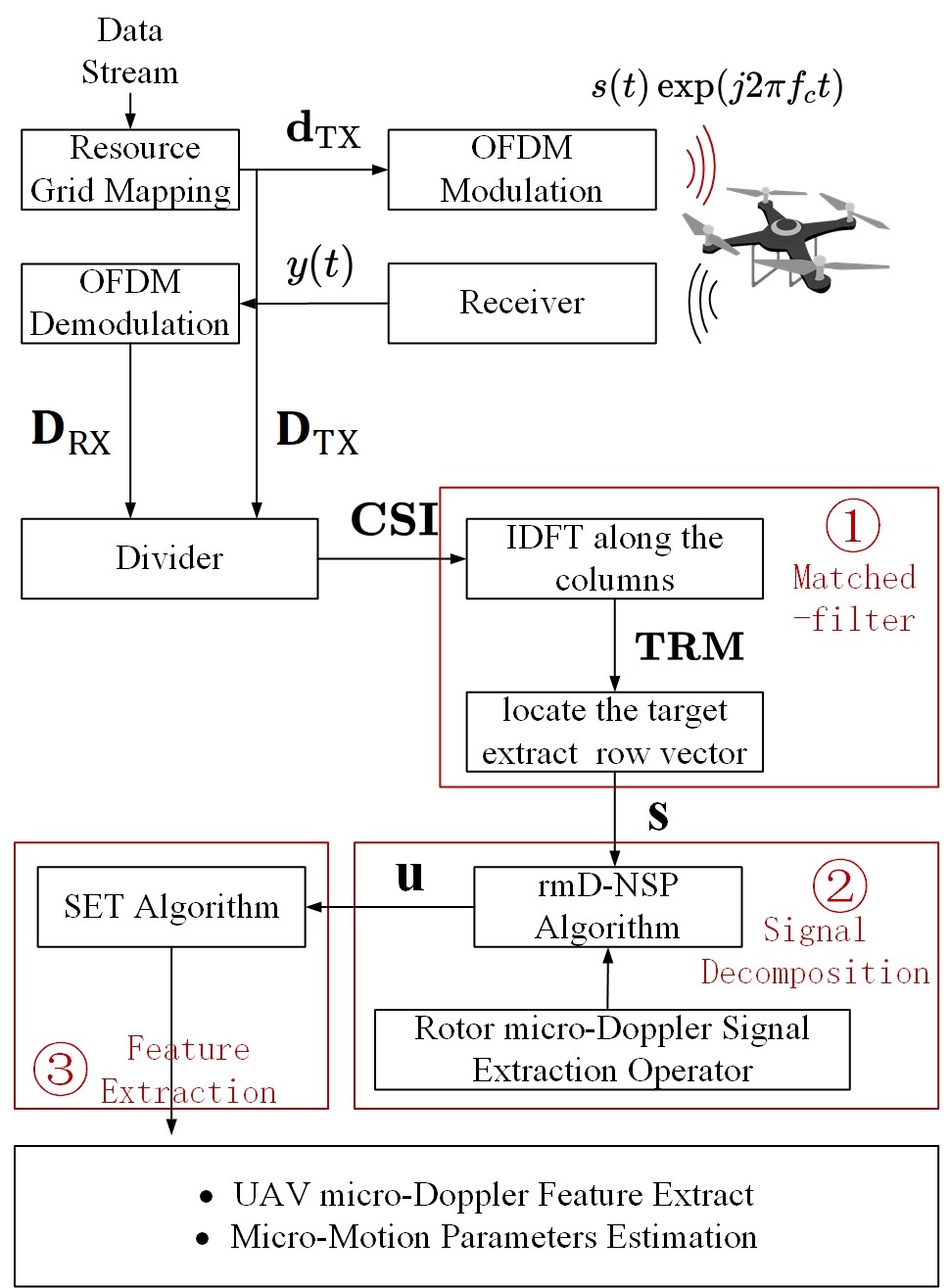}
    \caption{Overall procedure of UAV feature extraction.}
    \label{Fig:4}
\end{figure}

In this section, we present a method for extracting UAV micro-Doppler features using ISAC \ac{bs}. The overall process is illustrated in Fig.~\ref{Fig:4}. First, in Subsection~III-A, we introduce the matched filtering method to determine the position of the UAV and give the mixed signal model to be decomposed. \textcolor{black}{Then in Subsection~III-B, an operator-based signal decomposition method, denoted by rmD-NSP, is adopted to extract the rotor micro-Doppler signals from the mixed signals.} Finally, in Subsection~III-C, the feature extraction method, \textcolor{black}{denoted by \ac{set}, is applied to obtain the high resolution rotor features form the decomposed signals.}

\begin{figure*}[!t] 
    \centering 
    \includegraphics[width=0.85\textwidth]{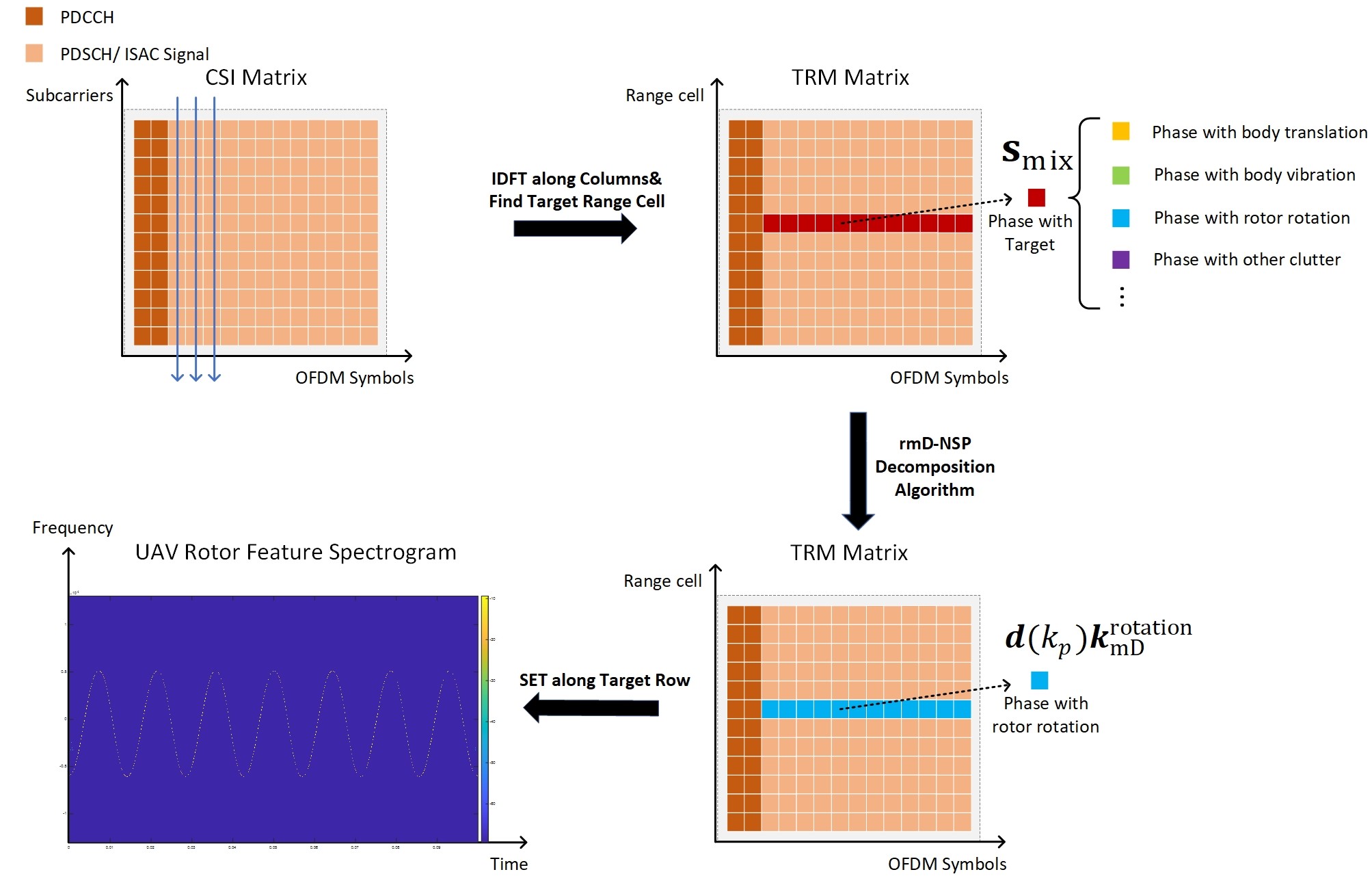} 
    \caption{The objective of the decomposition algorithm.}
    \label{fig:5} 
    
\end{figure*}
\subsection{Determine the UAV position}
According to (13) and (16), the range $R_0$ between the UAV and the \ac{bs} can be obtained by performing an \ac{idft} along the columns of the \ac{csi} matrix. The matrix after \ac{idft} processing is denoted as $\mathbf{TRM}$:
\begin{equation}
\begin{aligned}
&\mathbf{TRM}=\\
    &\left(\begin{array}{cccc}\boldsymbol{d}\left(1\right)\boldsymbol{k}_{D} \left(1\right) & \boldsymbol{d}\left(1\right)\boldsymbol{k}_{D} \left(2\right) & \cdots & \boldsymbol{d}\left(1\right)\boldsymbol{k}_{D} \left(M\right) \\\boldsymbol{d}\left(2\right)\boldsymbol{k}_{D} \left(1\right) & \boldsymbol{d}\left(2\right)\boldsymbol{k}_{D} \left(2\right) & \cdots & \boldsymbol{d}\left(2\right)\boldsymbol{k}_{D} \left(M\right) \\\vdots & \vdots & \ddots & \vdots \\\boldsymbol{d}\left(N\right)\boldsymbol{k}_{D} \left(1\right) & \boldsymbol{d}\left(N\right)\boldsymbol{k}_{D} \left(2\right) & \cdots & \boldsymbol{d}\left(N\right)\boldsymbol{k}_{D} \left(M\right)\end{array}\right),
\end{aligned}
\end{equation}
where $\mathbf{TRM} \in \mathbb{C}^{N \times M}$, and $\boldsymbol{d} \in \mathbb{C}^{N\times1}$ is given by,
\begin{equation}
\begin{aligned}\boldsymbol{d}\left(k\right)= & \operatorname{IDFT}\left[\boldsymbol{k}_{R}\left(n\right)\right]=\frac{1}{N} \sum_{n=1}^{N} \boldsymbol{k}_{R}\left(n\right) \exp \left(j \frac{2 \pi}{N} n k\right) \\= & \frac{1}{N} \sum_{n=1}^{N} \exp \left(-j 2 \pi  n \Delta f \frac{2 R_{0}}{c}\right) \exp \left(j \frac{2 \pi}{N} n k\right), \\& k=1, \ldots, N .
\end{aligned}    
\end{equation}
When the UAV is present, $\left |\boldsymbol{d}\right |$ will exhibit a peak. \textcolor{black}{Let $k_{p} \in \left\{1, 2, \ldots, N\right\}$ denote the position where the peak appears. Then, we extract the $k_{p}$-th row vector, denoted by $\boldsymbol{\mathrm{s}}_{\mathrm{uav}} \in \mathbb{C}^{1 \times M}$, from the $\mathbf{TRM}$,}
\begin{equation}
    \textcolor{black}{\boldsymbol{\mathrm{s}}_{\mathrm{uav}} =\mathbf{TRM}\left(k_{p},:\right) =  \boldsymbol{d}\left(k_p\right)\boldsymbol{k}_{D}.}
\end{equation}

In actual urban environments, the echoes contain not only the target but also significant strong interfering clutter. Considering this, the mixed signal model $\boldsymbol{\mathrm{s}}_{\text{mix}}\in \mathbb{C}^{M \times 1}$ to be decomposed is formulated as follows: 
    \begin{equation}
    \begin{aligned}
        \textcolor{black}{\boldsymbol{\mathrm{s}}_{\text{mix}} = \boldsymbol{\mathrm{s}}_{\mathrm{uav}}^{\mathrm{T}} +\boldsymbol{\eta}_{c},}
    \end{aligned}
    \end{equation}
where $\boldsymbol{\eta}_{c}\in \mathbb{C}^{M \times 1}$ represents the clutter and noise vector.

According to (14) and (20), \textcolor{black}{$\boldsymbol{\mathrm{s}}_{\text{mix}}$} contains the body Doppler information $\textcolor{black}{\boldsymbol{k}_{D}^{\text{trans}}}$, the body vibration micro-Doppler information \textcolor{black}{$\boldsymbol{k}_{mD}^{\text{vibration}}$}, the micro-Doppler information \textcolor{black}{$\boldsymbol{k}_{mD}^{\text{rotation}}$} caused by the rotation of the rotors and the clutter. Meanwhile, if \textcolor{black}{the} micro-Doppler feature extraction \textcolor{black}{processing} is directly performed on \textcolor{black}{$\boldsymbol{\mathrm{s}}_{\text{mix}}$}, the resulting time-frequency spectrogram not only contains the rotor micro-Doppler features but also includes significant interference information. The micro-Doppler signals generated by the body vibration overlap with the rotor micro-Doppler signals, affecting the feature extraction results~\cite{UAVvibration1}. Moreover, \textcolor{black}{the weak rotor micro-Doppler signals are obscured by strong interference signals such as clutter and echoes from the UAV body.}

\textcolor{black}{According to (6) and (15), the unique motion characteristics of the rotors modulate the echo in both amplitude and phase. Our approach is to leverage the distinctive amplitude-phase information of the rotor micro-Doppler signals $\boldsymbol{k}_{mD}^{\text{rotation}}$ to separate them from the mixed signal $\boldsymbol{\mathrm{s}}_{\text{mix}}$ using a signal decomposition method. Subsequently, feature extraction processing is performed on the decomposed signal.}
\subsection{Operator-based Signal Decomposition Algorithm}
    In Section~III-A, we \textcolor{black}{obtain a mixed signal $\boldsymbol{\mathrm{s}}_{\text{mix}}$ that includes the rotor micro-Doppler signals $\boldsymbol{d}\left(k_p\right)\boldsymbol{k}_{mD}^{\text{rotation}}$ along with various interference signals.} Our goal is to obtain \textcolor{black}{$\boldsymbol{d}\left(k_p\right)\boldsymbol{k}_{mD}^{\text{rotation}}$ from $\boldsymbol{\mathrm{s}}_{\text{mix}}$} to extract the UAV rotor micro-Doppler features as shown in Fig.~\ref{fig:5}. This is essentially a multi-component signal decomposition problem. Several multi-component signal decomposition algorithms have been studied, such as \ac{emd} and \ac{vmd}~\cite{huang1998EMD,VMD}. However, \textcolor{black}{\ac{emd} and \ac{vmd} do not perform signal decomposition based on the form of the target signal, making them unsuitable for decomposing mixed signal $\boldsymbol{\mathrm{s}}_{\text{mix}}$ to extract the micro-Doppler signals.}
    
    \ac{nsp} is an adaptive operator-based signal decomposition method that constructs adaptive operators to remove local narrowband signals in the zero space for separation~\cite{pengNSP,PengNSP1,AMFMNSP}. \textcolor{black}{The main appeal of this algorithm lies in its ability to design an adaptive operator based on the target signal model, enabling the extraction of the target signal from the mixed signal.} The algorithm implementation consists of two main steps: (i) Design a parameterized operator based on the form of the target signal. (ii) Use null-space tracking to adaptively adjust the operator parameters while simultaneously extracting the target signal. In this paper, the target signal is the rotor micro-Doppler signal \textcolor{black}{$\boldsymbol{d}\left(k_p\right)\boldsymbol{k}_{mD}^{\text{rotation}}$}. By designing the specific operator, we can achieve effective separation of the target signal.
    \subsubsection{Parameterized Operator Design.}
    Specifically, the NSP algorithm leverages \textcolor{black}{the} prior information about the \textcolor{black}{target signal}. It designs adaptive operators  based on the mathematical representation of the target signal, ensuring that the target signal falls into the zero space of the operator. 

By observing the mathematical representation of rotor micro-Doppler signal \textcolor{black}{$\boldsymbol{d}\left(k_p\right)\boldsymbol{k}_{mD}^{\text{rotation}}$} and according \textcolor{black}{to} (6), we can see that \textcolor{black}{$\boldsymbol{d}\left(k_p\right)\boldsymbol{k}_{mD}^{\text{rotation}}$} is composed of the product of an amplitude term and a phase term, where the phase term includes linearly time-varying Doppler information \textcolor{black}{$f_{D}\left(m\right)$} and non-linearly time-varying micro-Doppler information \textcolor{black}{$f_{mD}^{p}\left(m\right)$} in their frequency modulation. Furthermore, due to the dynamic \ac{rcs} characteristics caused by the high-speed rotation of the blades, rotor micro-Doppler signal also exhibits amplitude modulation \textcolor{black}{$\gamma^{rotor}_{p}\left(m\right)$} according to (6), where the modulation frequency is related to the rotational speed. Therefore, our idea is to utilize the unique amplitude-phase information of rotor signal to construct a operator $\mathcal{T}_{mD}$, ensuring that \textcolor{black}{$\boldsymbol{d}\left(k_p\right)\boldsymbol{k}_{mD}^{\text{rotation}}$} falls into the null space of this operator. 

In actual hardware processing, the signal is divided into real and imaginary parts for separate acquisition. \textcolor{black}{The operator $\mathcal{T}_{mD}$ thus effectively process on the real and imaginary parts of the signal. The operator processes both the real and imaginary components of the signal in an identical manner. Here, we take the real part as an example,}
\begin{equation}
    \textcolor{black}{\mathcal{T}_{mD}\left(\Re\left\{\boldsymbol{d}\left(k_p\right)\boldsymbol{k}_{mD}^{\text{rotation}}\left(m\right)\right\}\right) = 0.}
\end{equation}

According to (15), \textcolor{black}{the real part of rotor micro-Doppler siganl} $\textcolor{black}{\Re\left\{\boldsymbol{d}\left(k_p\right)\boldsymbol{k}_{mD}^{\text{rotation}}\left(m\right)\right\}}$ can be simplified to $ a\left(m\right) \cos \left(\phi_1\left(m\right)+\phi_2\left(m\right)\right)$, where \textcolor{black}{$a\left(m\right) = \Re\left\{\boldsymbol{d}\left(k_p\right)\gamma^{rotor}_{p}\left(m\right)\right\}$} contains amplitude modulation information, varying with changes in \ac{rcs}, \textcolor{black}{$\phi_1\left(m\right) = f_{mD}^{p}\left(m\right)$} represents micro-Doppler information, $\phi_2\left(m\right) = f_{D}\left(m\right)$ represents Doppler information.

Typically, for signals of this form, a second-order differential parameterized operator $\mathcal{T}_{mD}$ is chosen to extract the target signal~\cite{PengNSP1}:
\begin{equation}
\mathcal{T}_{mD} = \mathcal{D}_{2} 
+\mathbf{p}\left(m\right)\mathcal{D}_{1} +\mathbf{q}\left(m\right),
\end{equation}
where $\mathcal{D}_{2} $ is defined as the second-order difference operator and $\mathcal{D}_{1}$ is defined as the first-order difference operator. $\mathbf{p}\in \mathbb{R}^{M\times 1}$ and $\mathbf{q}\in \mathbb{R}^{M\times 1}$ are the parameters of the operator.

We substitute the simplified target signal expressions into (21),
\begin{equation}
\mathcal{T}_{mD}\left(a\left(m\right) \cos \left(\phi_1\left(m\right)+\phi_2\left(m\right)\right)\right)=0.
\end{equation}

 According to (23), the expressions for the parameters $\mathbf{p}\left(m\right)$ and $\mathbf{q}\left(m\right)$ in the operator $\mathcal{T}_{mD}$ can be derived. Due to the extremely small sampling interval $T_{s}$, the derivative of the sampled discrete-time sequence with respect to time closely approximates the derivative of the actual continuous-time form. Let $x^{\prime}\left(m\right) \triangleq\frac{\mathrm{d}x\left(t\right)}{\mathrm{d}t}\mid _{t=mT_{s}}$, $x^{\prime\prime}\left(m\right) \triangleq \frac{\mathrm{d}^{2}x\left(t\right)}{\mathrm{d}t^{2}}\mid _{t=mT_{s}}$, and $\bar{\phi}\left(m\right)=\phi_1{ }^{\prime}\left(m\right)+\phi_2{ }^{\prime}\left(m\right)$, the following equation is given:
\begin{equation}
\begin{aligned}
\left\{\begin{array}{l}-2 a^{\prime}\left(m\right)\bar{\phi}\left(m\right)-a\left(m\right)\left(\bar{\phi}{ }^{\prime}\left(m\right)-\bar{\phi}\left(m\right) \mathbf{p}\left(m\right)\right)=0, \\a^{\prime \prime}\left(m\right)+a^{\prime}\left(m\right) \mathbf{p}\left(m\right)+a\left(m\right)\left( \mathbf{q}\left(m\right)-\bar{\phi}^2\left(m\right)\right)=0.\end{array}\right.
\end{aligned}
\end{equation}

\textcolor{black}{Since $\phi_2\left(m\right)$ is a linear first-order function of $m$, $ \phi_{2}^{\prime\prime}\left(m\right)= 0$.} Then we obtain the expressions for $\mathbf{p}\left(m\right)$ and $\mathbf{q}\left(m\right)$:
\begin{equation}
\begin{aligned}
\left\{\begin{array}{l}\mathbf{p}\left(m\right)=-2 \frac{a^{\prime}\left(m\right)}{a\left(m\right)}-\frac{\phi_1^{\prime \prime}\left(m\right)}{\bar{\phi}\left(m\right)}, \\\mathbf{q}\left(m\right)=\bar{\phi}\left(m\right)^2+2\left(\frac{a^{\prime}\left(m\right)}{a\left(m\right)}\right)^2+\frac{a^{\prime}\left(m\right)}{a\left(m\right)} \frac{\phi_1^{\prime \prime}\left(m\right)}{\bar{\phi}\left(m\right)}-\frac{a^{\prime \prime}\left(m\right)}{a\left(m\right)},\end{array}\right.
\end{aligned}
\end{equation}
where $\left|\frac{a^{\prime}\left(m\right)}{a\left(m\right)}\right|$ is the instantaneous bandwidth of the signal. We assume that $\bar{\phi}\left(m\right)^2+2\left(\frac{a^{\prime}\left(m\right)}{a\left(m\right)}\right)^2+\frac{a^{\prime}\left(m\right)}{a\left(m\right)} \frac{\phi_1^{\prime \prime}\left(m\right)}{\bar{\phi}\left(m\right)} \gg \frac{a^{\prime \prime}\left(m\right)}{a\left(m\right)}$ \textcolor{black}{according to~\cite{AMFMNSP}.} Then we obtain the operator $\mathcal{T}_{mD}$,
\begin{equation}
\begin{aligned}
\mathcal{T}_{mD} =& \mathcal{D}_{2} +\left(-2 \frac{a^{\prime}\left(m\right)}{a\left(m\right)}-\frac{\phi_1^{\prime \prime}\left(m\right)}{\bar{\phi}\left(m\right)}\right) \mathcal{D}_{1}  \\+&\bar{\phi}\left(m\right)^2
+2\left(\frac{a^{\prime}\left(m\right)}{a\left(m\right)}\right)^2+\frac{a^{\prime}\left(m\right)}{a\left(m\right)} \frac{\phi_1^{\prime \prime}\left(m\right)}{\bar{\phi}\left(m\right)}.
\end{aligned}
\end{equation}

At this point, the real part of the rotor micro-Doppler signal is in the null space of the designed operator. The operator design is complete.

\subsubsection{Null Space Pursuit Algorithm}
According to the \ac{nsp} theory, after designing the operator $\mathcal{T}_{mD}$ specific to the real part of target signal, the following optimization problem can be solved to estimate the parameters of the operator, and separate the real part of the mixed signal $\boldsymbol{\mathrm{s}}_{\text{real}} = \Re\{\boldsymbol{\mathrm{s}}_{\text{mix}}\}\in \mathbb{R}^{M\times 1}$ into the target signal $\mathbf{u} = \boldsymbol{\mathrm{s}}_{\text{real}} - \mathbf{r}\in \mathbb{R}^{M\times 1}$ and the residual $\mathbf{r}\in \mathbb{R}^{M\times 1}$,
\begin{equation}
\begin{aligned} &\min _{\mathbf{r}\left ( m \right ) ,\mathbf{p}\left ( m \right ) ,\mathbf{q}\left ( m \right ), \lambda_{1}, \gamma}\left\{\left\|\mathcal{T}_{mD}\left(\boldsymbol{\mathrm{s}}_{\text{real}}\left ( m \right ) -\mathbf{r}\left ( m \right ) \right)\right\|^{2}\right. \\& \left.+\lambda_{1}\left(\| \mathbf{r}\left ( m \right ) \|^{2}+\gamma\|\boldsymbol{\mathrm{s}}_{\text{real}}\left ( m \right ) -\mathbf{r}\left ( m \right ) \|^{2}\right)\right.\\&
\left.+\lambda_{2}\left(\left\|\mathcal{D}_{2} \mathbf{q}\left(m\right)+\mathbf{p}\left(m\right)\right\|^{2}\right)\right\} ,\end{aligned}
\end{equation}
where $\lambda_{1}$ and $\lambda_{2}$ \textcolor{black}{are} the Lagrange multiplier. $\gamma$ is the leakage parameter determining the quantity of $\boldsymbol{\mathrm{s}}_{\text{real}}\left ( m \right ) -\mathbf{r}\left ( m \right ) $ preserved in the null space of $\mathcal{T}_{mD}$.


\textcolor{black}{We obtain the matrix form of corresponding augmented Lagrangian function as expressed in (27),}
\begin{equation}
    \begin{aligned}
    \mathcal{F}\left(\mathbf{r},\boldsymbol{\theta}, \lambda_{1},\gamma \right)= &\quad\left\|\mathbf{D}_{2}\left(\boldsymbol{\mathrm{s}}_{\text{real}}-\mathbf{r}\right)+\mathbf{A} \boldsymbol{\theta}\right\|^{2}\\ +&\lambda_{1}\left(\|\mathbf{r}\|^{2}+\gamma\|\boldsymbol{\mathrm{s}}_{\text{real}}-\mathbf{r}\|^{2}\right)\\
    +&\lambda_{2}\left\|\mathbf{M}_{2} \boldsymbol{\theta}\right\|^{2},
    \end{aligned}
\end{equation}
\textcolor{black}{ where $\boldsymbol{\theta}=\left[\mathbf{p}^{T}, \mathbf{q}^{T}\right]^{T}\in \mathbb{R}^{2M\times 1}$ contains the operator parameters. $\mathbf{A}=\left[\begin{array}{ll}\mathbf{A}_{\mathbf{D}_{1}(\boldsymbol{\mathrm{s}}_{\text{real}}-\mathbf{r})} & \mathbf{A}_{\boldsymbol{\mathrm{s}}_{\text{real}}-\mathbf{r}}\end{array}\right]\in \mathbb{R}^{M\times 2M}$. $\mathbf{M}_2 = \begin{bmatrix}\mathbf{D}_2 & \mathbf{E}\end{bmatrix}\in \mathbb{R}^{M\times 2M}$, where $\mathbf{E}\in \mathbb{R}^{M\times M}$ is the identity matrix.
}$\mathbf{D}_{2}\in \mathbb{R}^{M\times M}$ and $\mathbf{D}_{1}\in \mathbb{R}^{M\times M}$ can be written as follows, which represent the second-order differential matrice and the first-order differential matrice, respectively, 
\begin{equation}
\begin{aligned}
    \mathbf{D}_2=\left[\begin{array}{ccccc}-1 & 1 & 0 & \cdots & 0 \\1 & -2 & 1 & \cdots & 0 \\\vdots & \ddots & \ddots & \ddots & \vdots \\0 & \cdots & 1 & -2 & 1 \\0 & \cdots & 0 & 1 & -1\end{array}\right],
\end{aligned}
\end{equation}
\begin{equation}
\begin{aligned}
    \mathbf{D}_1=\left[\begin{array}{ccccccc}-1 & 1 & 0  & \cdots & 0 \\0 & -1 & 1 & \cdots & 0 \\\vdots & \ddots & \ddots  & \ddots & \vdots \\0 & \cdots   & 0 & -1 & 1 \\0 & \cdots & 0 & 0 & -1\end{array}\right].
\end{aligned}    
\end{equation}

 We can solve the above problems by using an iterative algorithm. By iteratively calculating the parameters $\mathbf{r}$, $\boldsymbol{\theta}$, $\lambda_{1}$ and $\gamma$, we can estimate the parameters of the operator $\mathcal{T}_{mD}$ and extract the target micro-Doppler signal $\mathbf{u}$. \textcolor{black}{In the $j$-th iteration,} the current residual signal is $\mathbf{r}^{j}$, the current leakage component is $\gamma^{j}$, and the current Lagrangian parameters is $\lambda_{1}^{j}$. The specific parameter update process and sequence are as follows:

(1) \textit{Iteration of} $\boldsymbol{\theta}^{j}$: The parameter $\boldsymbol{\theta}^{j} = \left[\mathbf{p}^{j^T}, \mathbf{q}^{j^T}\right]^{T}$ can be obtained by solving the following subproblem,
\begin{equation}
    \boldsymbol{\theta}^{j} = \arg \min _{\boldsymbol{\theta}}\quad\left\|\mathbf{D}_{2}(\boldsymbol{\mathrm{s}}_{\text{real}}-\mathbf{r}^{j})+\mathbf{A}^{j} \boldsymbol{\theta}\right\|^{2}+\lambda_{2}\left\|\mathbf{M}_{2} \boldsymbol{\theta}\right\|^{2},
\end{equation}
where $\mathbf{A}^{j} = \left[\begin{array}{ll}\mathbf{A}_{\mathbf{D}_{1}(\boldsymbol{\mathrm{s}}_{\text{real}}-\mathbf{r}^{j})} & \mathbf{A}_{\boldsymbol{\mathrm{s}}_{\text{real}}-\mathbf{r}^{j}}\end{array}\right]$. By setting the first derivatives of $\mathcal{F}\left(\mathbf{r},\boldsymbol{\theta}, \lambda_{1},\gamma \right)$ with respect to $\boldsymbol{\theta}$ to zero, we get
\begin{equation}
    \hat{\boldsymbol{\theta}}^{j}=-\left(\mathbf{A}^{j^T} \mathbf{A}^{j}+\lambda_{2} \mathbf{M}_{2}{ }^{T} \mathbf{M}_{2}\right)^{-1} \mathbf{A}^{j^T} \mathbf{D}_{\mathbf{2}}(\boldsymbol{\mathrm{s}}_{\text{real}}-\mathbf{r}^{j}).
\end{equation}

(2) \textit{Iteration of} $\lambda_{1}^{j+1}$: The parameter $\lambda_{1}^{j+1}$ is iteratively updated based on the \ac{nsp} theory~\cite{pengNSP} which can be calculated as follows:
\textcolor{black}{\begin{equation}
    \lambda_{1}^{j+1}=\frac{1}{1+\gamma^{j}} \frac{\boldsymbol{\mathrm{s}}_{\text{real}}^{T} \chi ^{j^T} \boldsymbol{\mathrm{s}}_{\text{real}}}{\boldsymbol{\mathrm{s}}_{\text{real}}^{T} \chi ^{j^T} \chi ^{j} \boldsymbol{\mathrm{s}}_{\text{real}}},
\end{equation}}where $\chi ^{j}=\left(\mathbf{T}_{mD}^{j^T} \mathbf{T}_{mD}^{j}+(1+\gamma^{j}) \lambda_{1}^{j} \mathbf{E}\right)^{-1}$ and $\mathbf{T}_{mD}^{j}=\mathbf{D}_{2}+\left[\mathbf{A}_{\mathbf{p}^{j}}\mathbf{A}_{\mathbf{q}^{j}}\right] \left[\mathbf{D}_{\mathbf{1}}{ }^{T} \mathbf{E}^{T}\right]^{T}$.

(3) \textit{Iteration of} $\mathbf{r}^{j+1}$: The parameter $\mathbf{r}^{j+1}$ can be obtained by solving the following subproblem
\begin{equation}
\begin{aligned}
    \mathbf{r}^{j+1} =& \arg \min _{\mathbf{r}}\quad\left\|\mathbf{D}_{2}(\boldsymbol{\mathrm{s}}_{\text{real}}-\mathbf{r})+\mathbf{A} \boldsymbol{\theta}^{j}\right\|^{2}\\    +&\lambda_{1}^{j+1}\left(\|\mathbf{r}\|^{2}+\gamma^{j}\|\boldsymbol{\mathrm{s}}_{\text{real}}-\mathbf{r}\|^{2}\right),
\end{aligned}
\end{equation}
by setting the first derivatives of $\mathcal{F}\left(\mathbf{r},\boldsymbol{\theta}, \lambda_{1},\gamma \right)$ with respect to $\mathbf{r}$ to zero, we get
\begin{equation}
\begin{aligned}
\mathbf{r}^{j+1}=&\left(\mathbf{T}_{mD}^{j^T} \mathbf{T}_{mD}^{j}+(1+\gamma^{j}) \lambda_{1}^{j+1} \mathbf{E}\right)^{-1}\\
    &\left(\mathbf{T}_{mD}^{j^T} \mathbf{T}_{mD}^{j} \boldsymbol{\mathrm{s}}_{\text{real}}+\lambda_{1}^{j+1} \gamma^{j} \boldsymbol{\mathrm{s}}_{\text{real}}\right).
\end{aligned}
\end{equation}

(4) \textit{Iteration of} $\gamma^{j+1}$: The parameter $\gamma^{j+1}$ can be iteratively updated according to the following equation~\cite{pengNSP}:
\begin{equation}
    \gamma^{j+1}=\frac{(\boldsymbol{\mathrm{s}}_{\text{real}}-\mathbf{r}^{j+1})^{T} \boldsymbol{\mathrm{s}}_{\text{real}}}{\|\boldsymbol{\mathrm{s}}_{\text{real}}-\mathbf{r}^{j+1}\|^{2}}-1.
\end{equation}

The value of the parameter $\lambda_{2}$ is less sensitive to the separation result which is set as a constant~\cite{pengNSP}.
The detailed procedures are illustrated in Algorithm 1.
\begin{algorithm}[htb] 
\caption{rmD-NSP to extract the micro-Doppler signal} 
\label{MD} 
\begin{algorithmic}[1] 
\REQUIRE ~~ 
Mixed signal $\boldsymbol{\mathrm{s}}_{\text{real}}$, stopping threhold $\varepsilon$ and the initial values of $\lambda_{2}$, $\lambda_{1}^{0} \text { and } \gamma^{0} \text {. }$
\STATE \textbf{Initialization:} Let $j=0$, $\mathbf{r}^{j} = 0$, $\lambda_{1}^{j}=\lambda_{1}^{0}$ and $\gamma^{j}=\gamma^{0}$ 
\REPEAT 
\STATE $\boldsymbol{\theta}^{j} \leftarrow \mathbf{r}^{j},\mathbf{A}^{j}$ according to (32). \
\STATE $\lambda_{1}^{j+1} \leftarrow \lambda_{1}^{j}, \gamma^{j}, \boldsymbol{\theta}^{j}$ according to (33).\
\STATE $\mathbf{r}^{j+1} \leftarrow \boldsymbol{\theta}^{j}, \gamma^{j}, \lambda_{1}^{j+1}$ according to (35).\
\STATE $\gamma^{j+1} \leftarrow \mathbf{r}^{j+1}$ according to (36), and set $j=j+1$ .
\UNTIL {$\left\|\mathbf{r}^{j+1}-\mathbf{r}^{j}\right\|^{2}> \varepsilon \left\|\boldsymbol{\mathrm{s}}_{\text{real}}\right\|^{2}$}
\ENSURE ~~\\ 
The extracted rotor micro-Doppler signal $\hat{\mathbf{u}}=\left(1+\gamma^{j}\right)\left(\boldsymbol{\mathrm{s}}_{\text{real}}-\mathbf{r}^{j}\right)$.
\end{algorithmic}
\end{algorithm}

\subsection{High-Resolution Feature Extraction Algorithm}
After applying the rmD-NSP decomposition algorithm from the previous section, we have now extracted the rotor micro-Doppler component $\boldsymbol{d}\left(k_p\right)\boldsymbol{k}_{mD}^{\text{rotation}}$ from $\boldsymbol{\mathrm{s}}_{\text{mix}}$. Next, we extract the \textcolor{black}{micro-Doppler} features of this signal. 
SET is a new time-frequency analysis method that can generate more energy-concentrated spectrogram results\cite{SET}. SET extracts time-frequency coefficients near the \textcolor{black}{\ac{if}} while discarding the remaining time-frequency coefficients, resulting in better performance for multi-component non-stationary signals. Therefore, we choose this algorithm as the method for subsequent feature extraction.

The main steps of this algorithm are:

(1) The decomposed signal $\mathbf{u}$ after rmD-NSP algorithm retains most of the rotor micro-Doppler signals. Perform modified STFT on this signal:
\begin{equation}
\begin{aligned}
S_{\mathbf{u}}\left(\eta, m\right) &= \sum_{\xi^{\prime}=-\infty}^{+\infty} \mathbf{u}\left[\xi^{\prime} + m\right] g^*\left[\xi^{\prime}\right] e^{-2\pi j \eta \xi^{\prime}}\\&=\sum_{\xi=-\infty}^{+\infty} \mathbf{u}\left[\xi\right] g\left[\xi-m\right] e^{-2\pi j \eta \left(\xi-m\right)}\\& =M_{\mathbf{u}}\left(\eta, m\right) e^{2 j \pi \Phi_{\mathbf{u}}\left(\eta, m\right)},
\end{aligned}
\end{equation}
where $g\left(m\right)$ is a finite-length window function, $M_{\mathbf{u}}\left(\eta, m\right)$ is the magnitude of STFT results, and $\Phi_{\mathbf{u}}\left(\eta, m\right)$ is the phase.

(2) Based on the modified STFT result $S_{\mathbf{u}}\left(\eta, m\right)$, we calculate the instantaneous frequency $\hat{\omega}_{\mathbf{u}}\left(\eta, m\right)$ of $\mathbf{u}\left(m\right)$ as follows:
\begin{equation}
    \hat{\omega}_{\mathbf{u}}\left(\eta, m\right)=\left\{\begin{array}{ll}\Re\left\{\frac{\partial_{m} S_{\mathbf{u}}\left(\eta, m\right)}{2 j \pi S_{\mathbf{u}}\left(\eta, m\right)}\right\}, & \left|S_{\mathbf{u}}\left(\eta, m\right)\right|>\gamma, \\\infty, & \left|S_{\mathbf{u}}\left(\eta, m\right)\right| \leq \gamma,\end{array}\right.
\end{equation}
where $\partial_{m} S_{\mathbf{u}}\left(\eta, m\right)\triangleq\frac{S_{\mathbf{u}}\left(\eta, m+1\right) - S_{\mathbf{u}}\left(\eta, m\right)}{T_s}\approx \frac{\mathrm{d}S_{\mathbf{u}}\left(\eta, t\right)}{\mathrm{d}t} \bigg|_{t = mT_s}$. $\gamma > 0$ is to eliminate the unstable phenomenon or the influence of the noises.

(3) Perform energy extraction to obtain the SET result:
\begin{equation}
\begin{aligned}
\mathrm{SET}_{\mathbf{u}}\left(\eta, m\right)&=S_{\mathbf{u}}\left(\eta, m\right) \delta\left(\eta-\hat{\omega}_{\mathbf{u}}\left(\eta, m\right)\right)\\
&=\left\{\begin{array}{ll}S_{\mathbf{u}}\left(\eta, m\right), & \eta=\hat{\omega}_{\mathbf{u}}\left(\eta, m\right) \\0, & \eta \ne \hat{\omega}_{\mathbf{u}}\left(\eta, m\right),\end{array}\right.
\end{aligned}
\end{equation}
where $\delta\left(\eta-\hat{\omega}_{\mathbf{u}}\left(\eta, m\right)\right)$ is used to gather the STFT coefficients that have the same frequency to where they should appear.

\begin{figure}[h]
    \centering
    
    \subfloat[]{\includegraphics[width=0.88\linewidth]{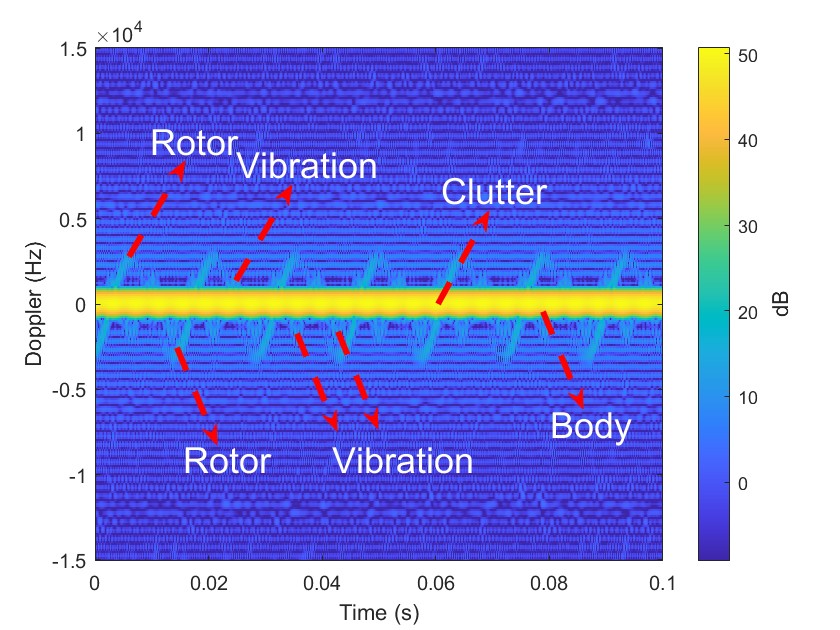}%
    }
    \hfil
    \subfloat[]{\includegraphics[width=0.88\linewidth]{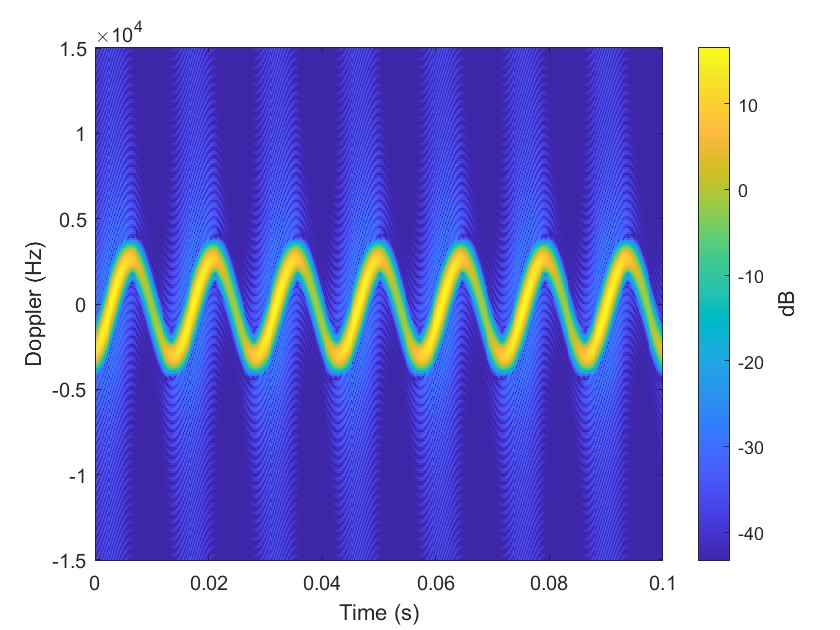}%
    }
    \caption{UAV Spectrogram feature: (a) Spectrogram of micro-Doppler feature extraction from raw UAV echo. (b) Theoretical spectrogram of micro-Doppler feature extraction from UAV rotor.}
    \label{Fig:6}
\end{figure}

\begin{figure*}[!t]
\centering
\subfloat[rmD-NSP first decomposition]{\includegraphics[width=2.3in]{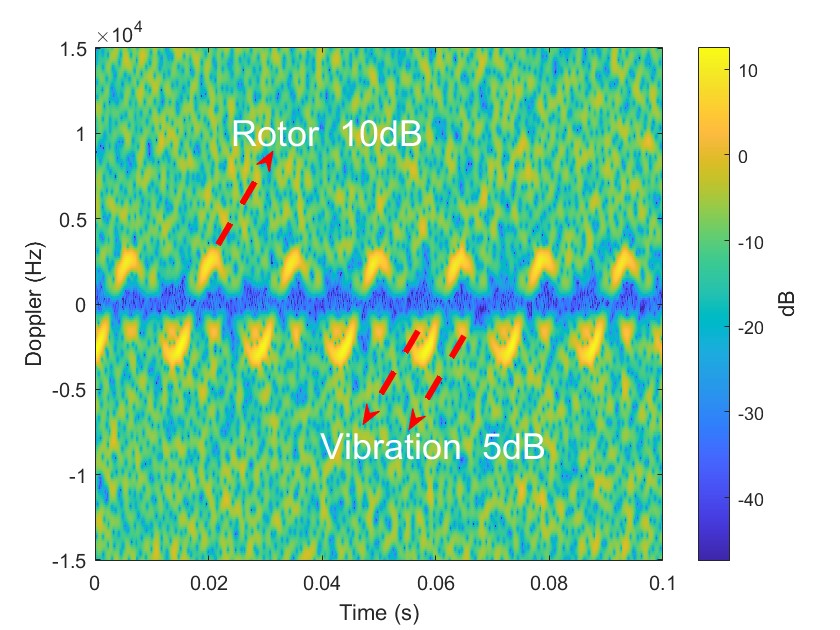}%
}
\hfil
\subfloat[rmD-NSP final decomposition]{\includegraphics[width=2.3in]{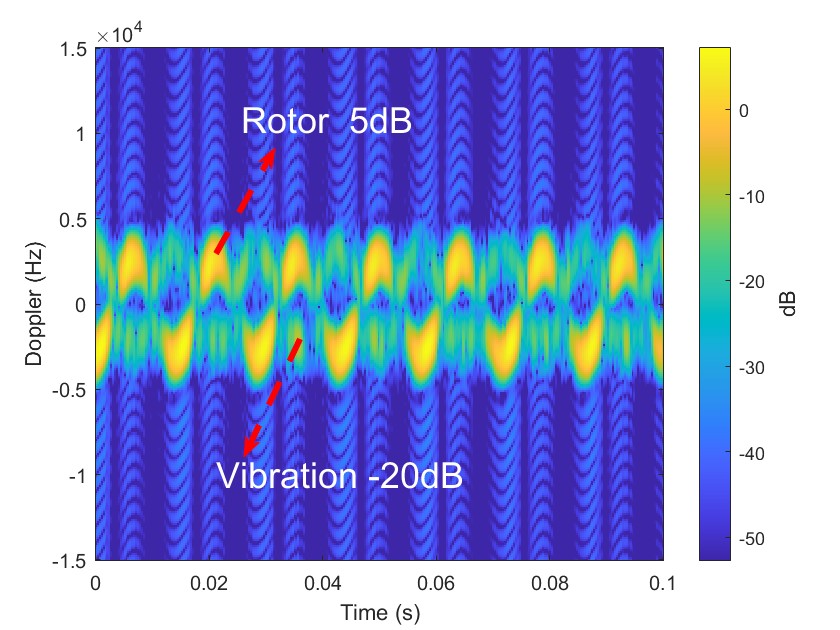}%
}
\hfil
\subfloat[SET after rmD-NSP final decomposition]{\includegraphics[width=2.3in]{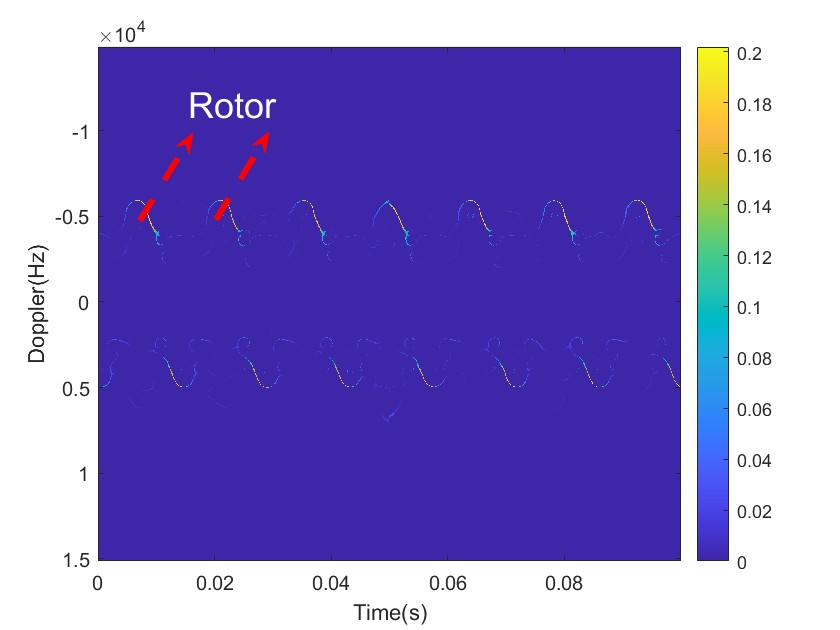}%
}
\hfil
\subfloat[AM-FM NSP first decomposition]{\includegraphics[width=2.3in]{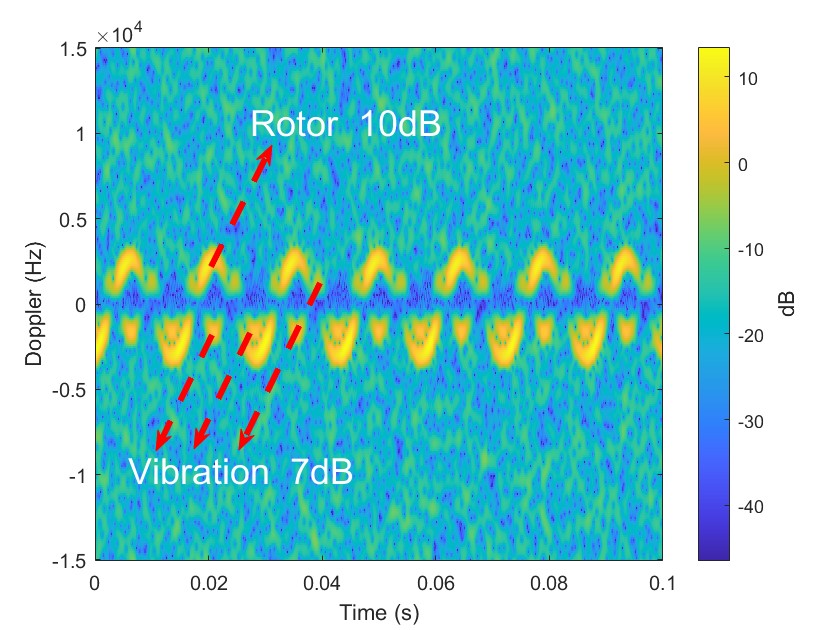}%
}
\hfil
\subfloat[AM-FM NSP final decomposition]{\includegraphics[width=2.3in]{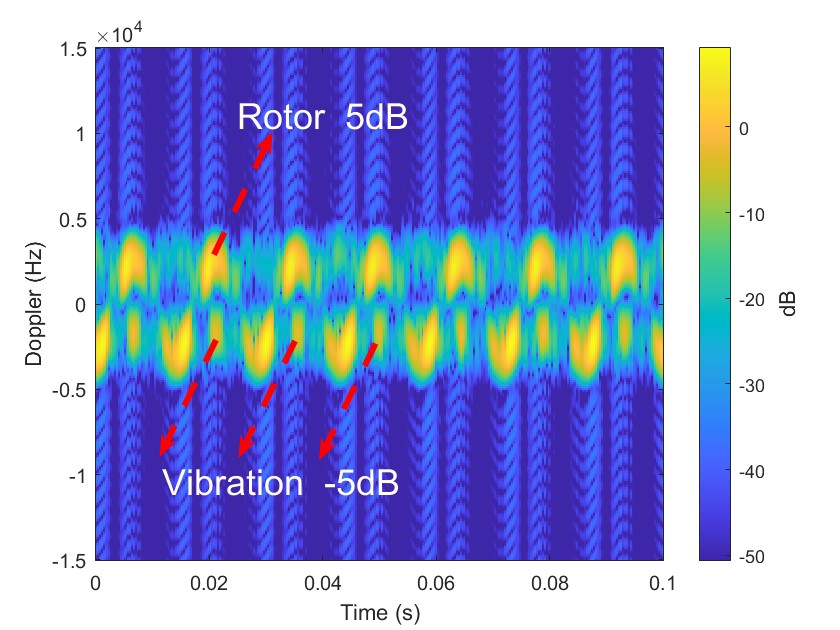}%
}
\hfil
\subfloat[SET after AM-FM NSP final decomposition]{\includegraphics[width=2.3in]{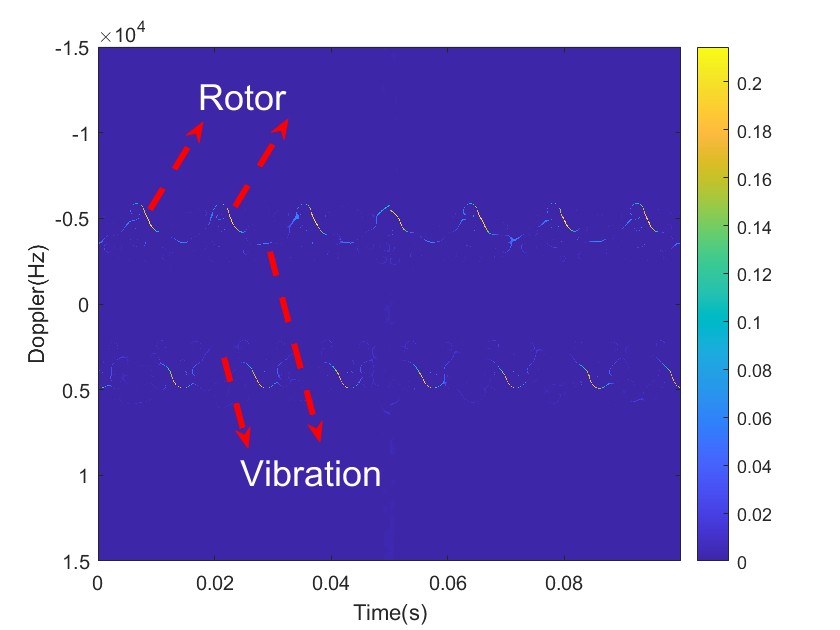}%
}
\hfil
\subfloat[NSP first decomposition]{\includegraphics[width=2.3in]{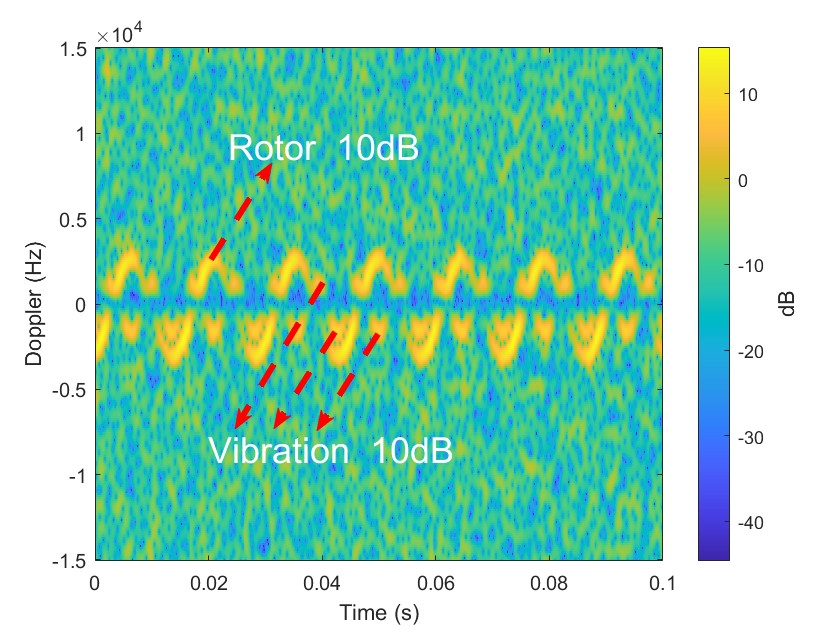}%
}
\hfil
\subfloat[NSP final decomposition]{\includegraphics[width=2.3in]{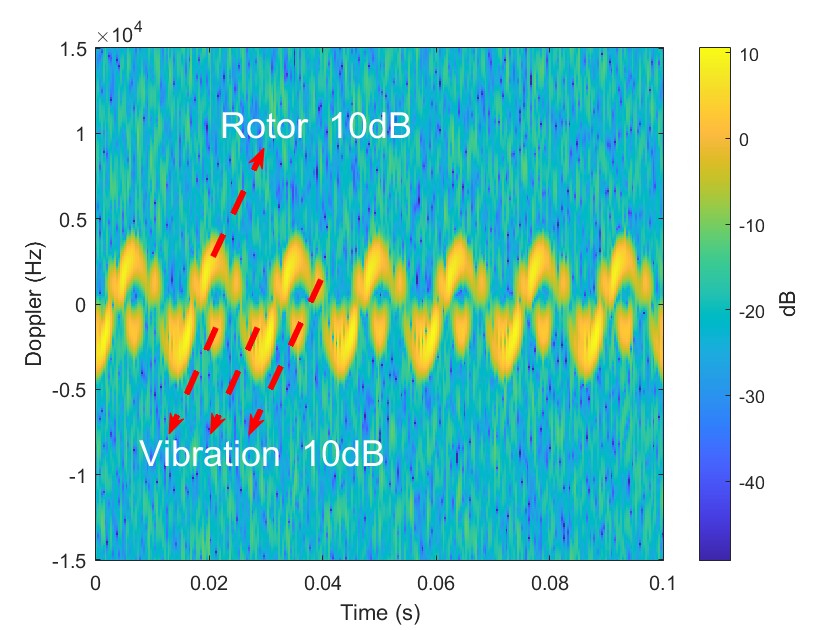}%
}
\hfil
\subfloat[SET after NSP final decomposition]{\includegraphics[width=2.3in]{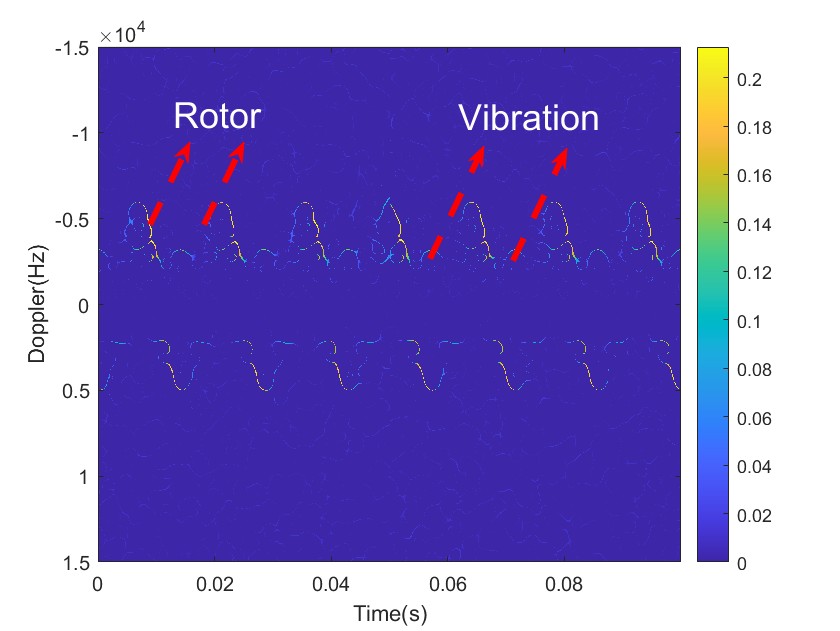}%
}
\hfil
\caption{Micro-Doppler feature extraction simulated by different methods. First row: (a) and (b) Spectrogram of the signals extracted using rmD-NSP, (c) spectrogram of the extracted signal using rmD-NSP after applying the SET algorithm. Second row: (d) and (e) Spectrogram of the signals extracted using AM-FM~NSP~\cite{AMFMNSP}, (f) spectrogram of the extracted signal using AM-FM~NSP after applying the SET algorithm. Last row: (g) and (h) Spectrogram of the signals extracted using NSP~\cite{pengNSP}, (i) spectrogram of the extracted signal using NSP after applying the SET algorithm.}
\label{Fig:7}
\end{figure*}

\section{SIMULATION RESULTS
}
\vspace{11pt}
This section verifies the ability to extract \ac{uav} features using ISAC signal through software simulation. 
The simulation scenario is based on a DJI M300 RTK UAV and a Sub-6GHz \ac{bs}, with the relative position of the \ac{bs} and the UAV shown in Fig.~\ref{Fig:2}. The UAV is modeled as a set of scatterers performs uniform linear motion along the radial direction of the \ac{bs}. 
The scatterers on the rotor exhibit both translational and rotational motions, as shown in Fig.~\ref{Fig:2}. The scatterers on the body exhibit both translational and vibrational motions, as shown in Fig.~\ref{Fig:3}. The parameters for rotation and vibration are provided in Table I.
\begin{table}[!h]
	\caption{\textcolor{black}{SIMULATION PARAMETERS}}
	\begin{center}
		\begin{tabular}{l l l}
			\hline
			\hline
			
			{Parameters} & {Symbol} &{Value} \\
			
			\hline
			ISAC siganl &  &$\text{PDSCH}$\\
			Carrier frequency & $f_c$ &\SI{3.5}{GHz} \\
			Subcarrier spacing & $\Delta f$ &\SI{30}{kHz}\\
			Bandwidth & $B$ &\SI{100}{MHz}\\
			TDD frame structure & &DDDSU\\
                Duration of a OFDM symbol& $T_s$ &\SI{36.6}{\mu\mathrm{s}} \\
                Coherent processing interval & &\SI{0.1}{s} \\
                Number of downlink sensing symbols & $M$ &1840 \\
                Number of subcarriers & $N$ &3276 \\
			Total transmission power &$P_t$ &\SI{28}{dBm} \\
                Antenna gains &$G_t,G_r$ &\SI{18}{dB} \\
                Distance between the UAV and the BS& $R_0$&\SI{50}{m} \\
                Azimuth angle & $\alpha$&\SI{0}{^{\circ}} \\
			Elevation angle & $\theta$&\SI{30}{^{\circ}} \\
			Velocity of the UAV & $v$ &\SI{5}{m/s} \\
			UAV rotation speed & $f_r$ &\SI{80}{r/s} \\
			Length of blade & L&\SI{0.5}{m}\\
                Azimuth angle of vibration & $\alpha_q$&\SI{10}{^{\circ}} \\
			Elevation angle of vibration & $\theta_q$&\SI{0}{^{\circ}} \\
                Vibration frequency& $f_v$&\SI{100}{Hz} \\
                Vibration amplitude& $D_v$&\SI{0.05}{m} \\
                RCS of the UAV body& $\sigma^{\text{body}}$&\SI{0.1}{m^2} \\
			\hline
			\hline
		\end{tabular}
	\end{center}
	\label{simu_para}
\end{table}
Furthermore, these scatterers also have different scattering intensities, and $\sigma^{\text{body}}$ is set to $\SI{0.1}{m^2}$, representing the \ac{rcs} of the \ac{uav} body, which does not change with time over a short \ac{cpi}. The RCS of the rotor blades can be assigned to the scatterers at the blade tips, denoted as $\sigma^{\text{rotor}}_{p}(t)$. The DJI M300 RTK is made of carbon fiber. According to~\cite{rcs}, we set $I$ in (6) and (7) to 6, and the specific coefficients $a_i,b_i$ and $c_i$ associated with the carbon fiber are given by
\begin{equation}
    \begin{aligned}\mathbf{a}= & {[1.133,0.425,0.7121,-0.1588,0.1046,0.0027] }, \\\mathbf{b}= & {[356.8,1445,608,1946,2236,3513] }, \\\mathbf{c}= & {[-0.1997,-2.464,1.695,1.319,-0.1277,} \\& -0.2433].\end{aligned}
\end{equation}
Due to the high-speed rotation of the UAV blades, $\sigma^{\text{rotor}}_{p}(t)$ changes rapidly within the CPI. This causes the echo intensity from the UAV rotor blades to vary rapidly over time.

The scattering intensity $\gamma(t)$ is ultimately expressed using the received power in the radar equation~\cite{radarEQ}, where $P_t$ is the transmit power, set to $\SI{28}{dBm}$~\cite{3gpp2019nr}. $G_t$ and $G_r$ are the transmitting and receiving antenna gains, which are set to $\SI{18}{dB}$, and $\lambda$ is the wavelength. $\sigma(t)$ is related to the type of scatterer. For scatterers on the UAV body, this term can be approximated as $\sigma(t) \approx \sigma^{\text{body}}$, and for scatterers on the rotor blades, this term can be approximated as $\sigma(t) \approx \sigma^{\text{rotor}}_{p}(t)$. $L_s$ is the system loss and can be ignored~\cite{radarEQ} and $L_a (R_0)$ is the path loss,
\begin{equation}
\gamma(t) = \frac{P_t G_t G_r \lambda^2 \sigma(t)}{(4 \pi)^3 R^4 L_s L_a (R_0)}.
\end{equation}

Let $\delta_{r}$ represent the system's range resolution, which is defined as follows:
\begin{equation}
    \delta_{r}=\frac{c}{2B},
\end{equation}
where $c$ represents the speed of light and $B$ represents the system bandwidth. With a system bandwidth of $\SI{100}{MHz}$, the range resolution $\delta_{r}$ is $\SI{1.5}{m}$. In this case, all scattering points on the UAV fall within a single range cell, and the rotation of the rotor does not span multiple distance units. By extracting the range cell where the target is located and performing time-frequency analysis, the spectrogram of the UAV can be extracted as shown in Fig.~\ref{Fig:6} (a). The rotor micro-Doppler siganls have a low intensity, and due to the presence of vibration components, the rotor micro-Doppler features are not clear. For comparison, Fig.~\ref{Fig:6} (b) shows the micro-Doppler feature with only one blade.
\begin{figure}[h]
    \centering
    \includegraphics[width=0.88\linewidth]{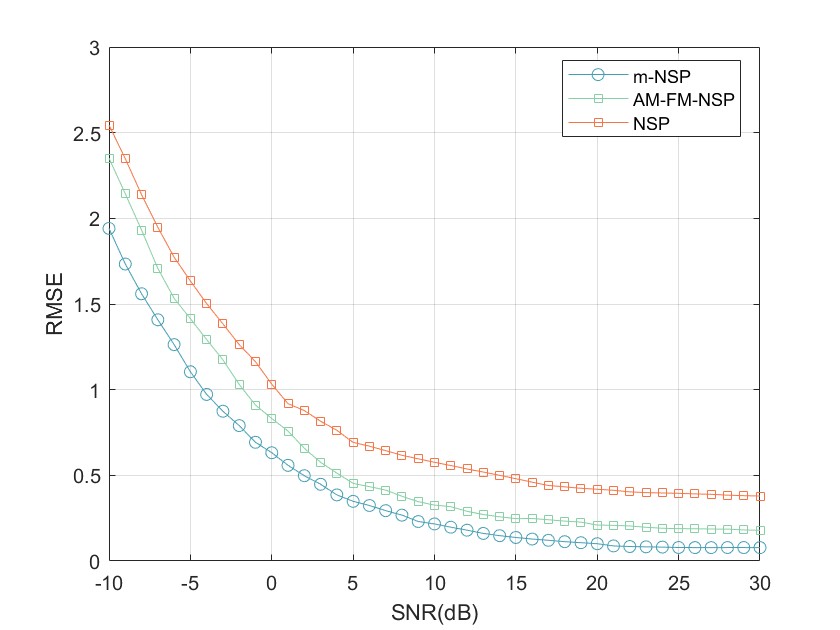}
    \caption{RMSE under different SNRs for the different decomposition algorithm.}
    \label{Fig:8}
\end{figure}

In order to extract the rotor micro-Doppler signals from the raw UAV echo, which contains numerous scatterers, and to obtain high-resolution rotor micro-Doppler characteristics, we apply three operator-based signal decomposition algorithms\textemdash NSP~\cite{pengNSP}, AM-FM NSP~\cite{AMFMNSP}, and the proposed rmD-NSP algorithm\textemdash on the mixed signal, i.e., $\boldsymbol{\mathrm{s}}_{\text{mix}}$ described in (20). Subsequently, we use the SET method to extract the micro-Doppler feature from the decomposed signal. The results are shown in Fig.~\ref{Fig:7}. The first row of Fig.~\ref{Fig:7} shows the results of decomposition and time-frequency feature enhancement using the rmD-NSP algorithm and SET. Fig.~\ref{Fig:7} (a) displays the results after the first decomposition, where the translation component of the UAV body and the stationary clutter are eliminated, but some vibration components have the same energy level as the rotor components, both being $\SI{5}{dB}$. Fig.~\ref{Fig:7} (b) shows the final decomposition result of rmD-NSP. The rotor component exhibits a high energy level, reaching up to $\SI{5}{dB}$, while the vibration component is only $\SI{-20}{dB}$. The rotor micro-Doppler characteristics are fully revealed from the environment, although the time-frequency ridge remains relatively coarse. Fig.~\ref{Fig:7} (c) shows the result of applying the SET to the final decomposition output of rmD-NSP, yielding a more refined time-frequency spectrogram. The second row of Fig.~\ref{Fig:7} shows the results of AM-FM NSP algorithm and SET algorithm. As shown in Fig.~\ref{Fig:7} (d), the first decomposition using the AM-FM NSP can still eliminate the UAV body Doppler and clutter components. Similar to rmD-NSP, it also retains some vibration components. Fig.~\ref{Fig:7} (e) shows the final decomposition result of AM-FM NSP, where it can be seen that some high-energy vibration components are still retained, reaching up to $\SI{-5}{dB}$, with energy levels close to those of the rotor. The last row of Fig.~\ref{Fig:7} shows the results of NSP and SET operations. From Fig.~\ref{Fig:7} (g) and Fig.~\ref{Fig:7} (h), it can be observed that the NSP algorithm consistently fails to eliminate the vibration components. The final decomposition results still retain all the vibration components, with energy levels as high as $\SI{10}{dB}$, matching that of the rotor. At this point, the time-frequency features on the spectrogram can no longer accurately represent the rotor's micro-Doppler features.

The reason for this issue is the difference in operators. First, the NSP operator only considers the frequency modulation component and has difficulty capturing the amplitude and frequency modulation components, such as the rotor's micro-Doppler signal. Therefore, during decomposition, it treats the vibration-induced micro-Doppler signal and the rotor-induced micro-Doppler signal as the same type of signal, resulting in residual vibration interference after decomposition. Both AM-FM NSP and rmD-NSP consider amplitude and frequency modulation components. However, rmD-NSP takes into account the characteristics of the rotor's micro-Doppler signal, with the frequency modulation component consisting of a linear Doppler term and a nonlinear micro-Doppler term according to (15). By utilizing this prior information, the algorithm can capture the differences between the vibration components and rotor components during the iterative decomposition process, resulting in better decomposition performance.

\begin{figure}[h]
    \centering
\includegraphics[width=0.8\linewidth]{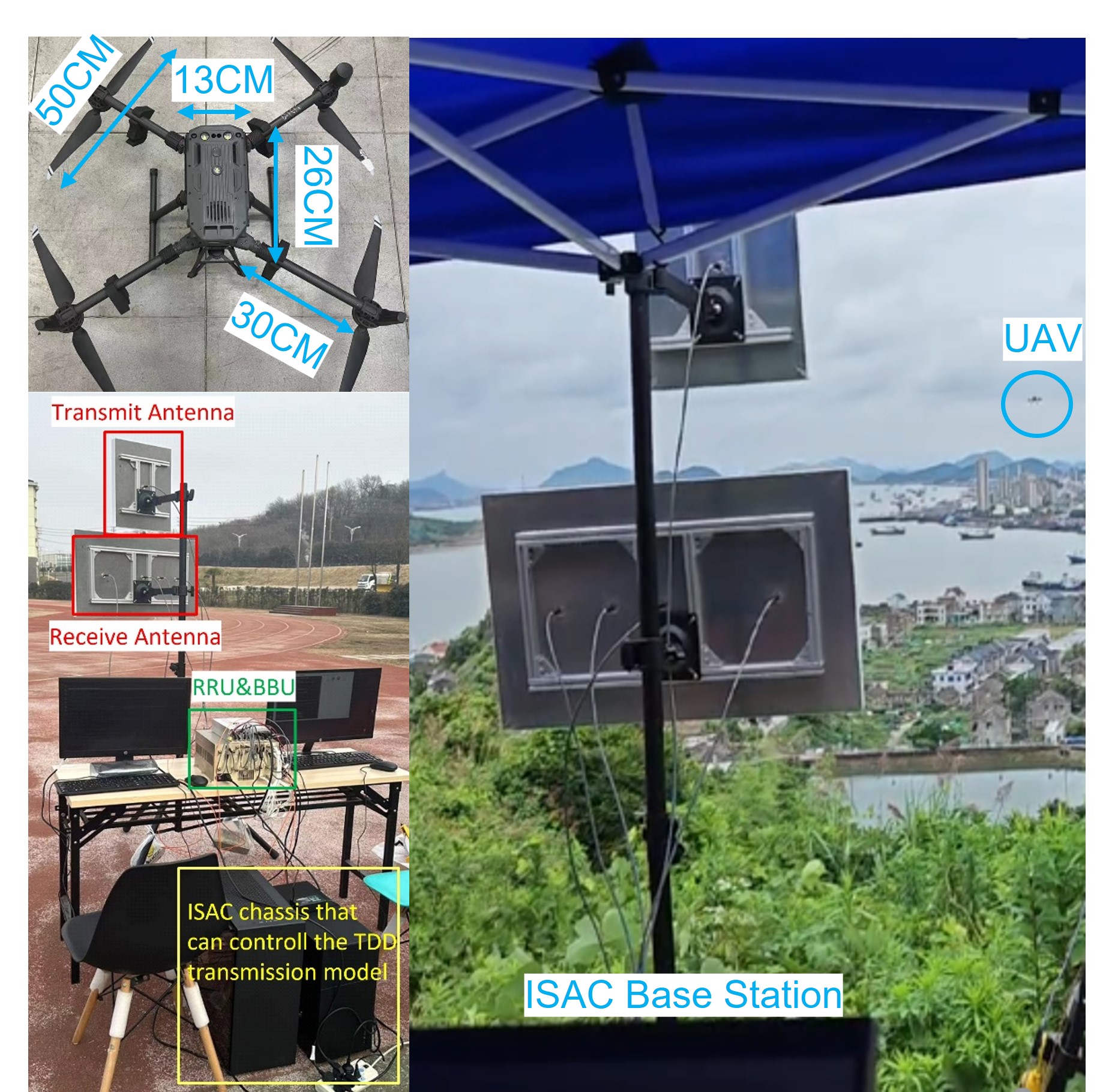}
    \caption{UAV micro-Doppler extraction test scenario based on ISAC hardware testbed.}
    \label{Fig:9}
\end{figure}
Further, we evaluate the decomposition performance of NSP, AM-FM NSP, and rmD-NSP algorithms on rotor micro-Doppler signals in a noisy environment. In this simulation, the \ac{mse} is used to assess the performance of NSP, AM-FM NSP, and rmD-NSP under noisy conditions. The \ac{snr} of the input signal increases in 1 dB increments from $\SI{-10}{dB}$ to $\SI{30}{dB}$. Fig.~\ref{Fig:8} presents the RMSE of the three algorithms in decomposing rotor micro-Doppler signals. It can be observed that under a $\SI{30}{dB}$ \ac{snr} condition, the rmD-NSP algorithm performs better, achieving the smallest RMSE of only 0.07, compared to AM-FM NSP at 0.17 and NSP at 0.38. At a $\SI{5}{dB}$ \ac{snr}, the rmD-NSP algorithm still performs well with an RMSE of 0.34, compared to the other two algorithms.
\begin{equation}
\mathrm{SNR} = \frac{\sum_{m=1}^{M}  \left |\boldsymbol{\mathrm{s}}_{\mathrm{uav}}\left(m\right)\right |^2 }{M\sigma_{n}^{2} },
\end{equation}

\section{HARDWARE TESTBED AND RESULTS
}
\vspace{11pt}

The proposed algorithm is verified on the Sub-6G ISAC hardware testbed. The ISAC system is based on the 5G NR protocol and consists of UL and DL baseband signal processing, intermediate frequency modulation, $\SI{3.5}{GHz}$ RF modulation, and antenna modules. The parameters for the Sub-6G ISAC hardware testbed are exactly the same as those listed in Table I. We utilize the proposed frame structure for sensing and communication functions. The test location is chosen in Ningbo, China, and the test scenario is shown in Fig.~\ref{Fig:9}.
\begin{figure}[h]
    \centering
    
    \subfloat[]{\includegraphics[width=0.83\linewidth]{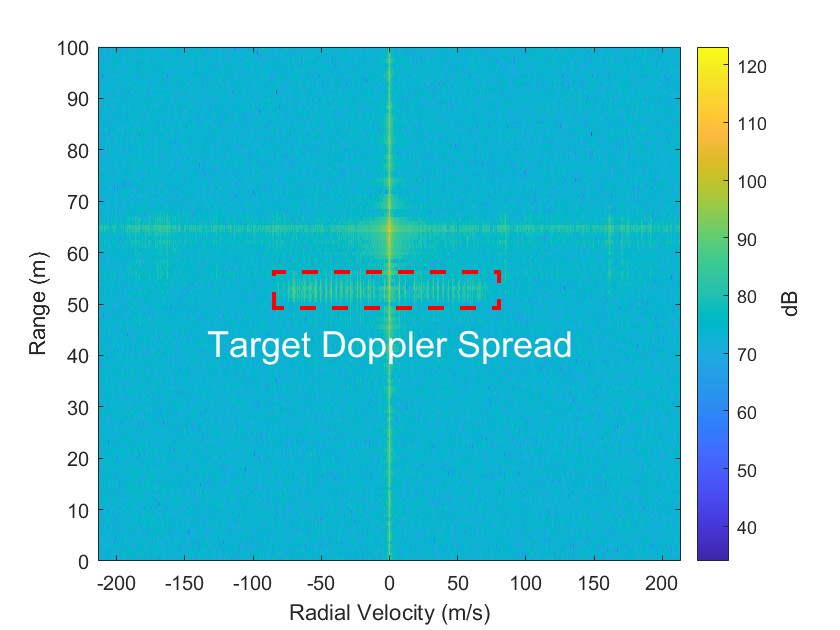}%
    }
    \hfil
    \subfloat[]{\includegraphics[width=0.83\linewidth]{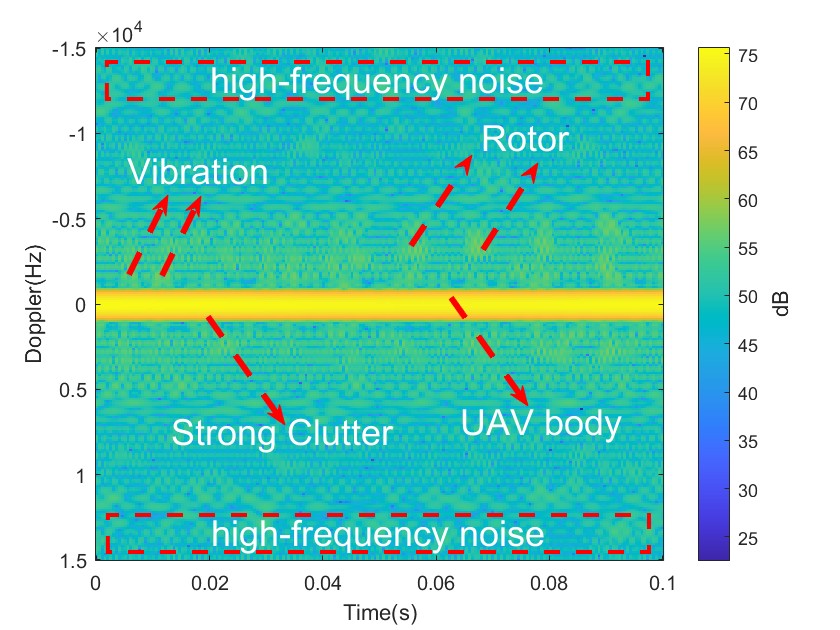}%
    }
    \caption{Hardware testbed results: (a) Range-Doppler map of UAV. (b) Spectrogram of raw UAV echo.}
    \label{Fig:10}
\end{figure}

\begin{figure*}[!t]
\centering
\subfloat[]{\includegraphics[width=2.33in]{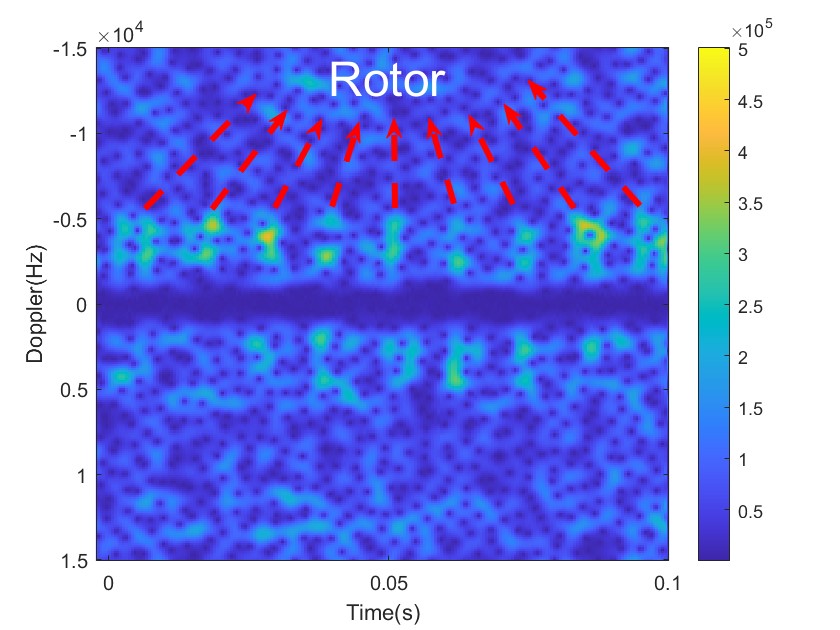}%
}
\hfil
\subfloat[]{\includegraphics[width=2.33in]{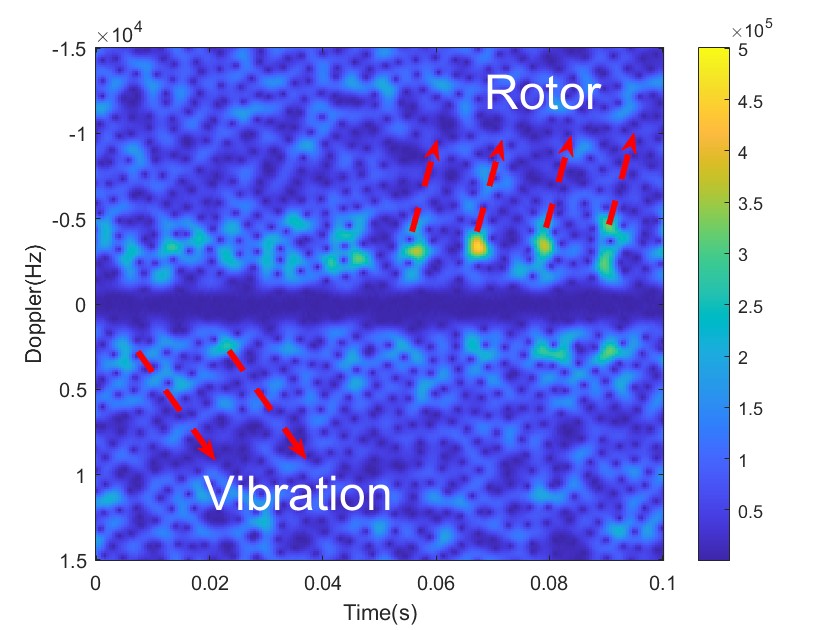}%
}
\hfil
\subfloat[]{\includegraphics[width=2.33in]{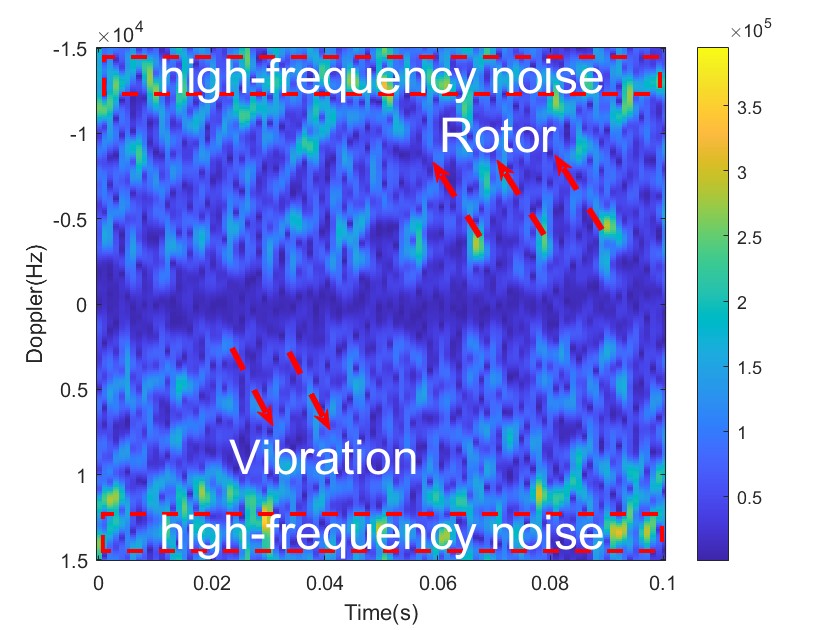}%
}
\hfil
\caption{Hardware testbed implementation results of the proposed algorithm: (a) Spectrogram after decomposition using the proposed rmD-NSP algorithm. (b) Spectrogram after decomposition using the AM-FM~NSP algorithm~\cite{AMFMNSP}. (c) Spectrogram after decomposition using the NSP algorithm~\cite{pengNSP}.}
\label{Fig:11}
\end{figure*}

The DJI M300 RTK UAV rotor length $L$ is $\SI{0.5}{m}$, and the body area is $\SI{0.26}{m} \times \SI{0.13}{m}$. We controlled the UAV to fly uniformly along the radial direction of the BS at a speed of $\SI{5}{m/s}$, while the ISAC BS collected the echoes to obtain the target's CSI information. Performing 2D-FFT~\cite{Sturm} on the CSI can obtain the range and Doppler information of the UAV, as shown in Fig.~\ref{Fig:10} (a). It can be seen that the UAV is not a point target on the range-Doppler map. At a distance of $\SI{53}{m}$, it exhibits a Doppler spread of $\SI{\pm 90}{m/s}$ along the velocity axis, which is caused by the rotation of the rotor and the vibration of the UAV body. Fig.~\ref{Fig:10} (b) shows the result of performing a \ac{stft} in the target range cell. It can be seen that there are many scattering points in the time-frequency spectrogram. Clutter and scatterers from the UAV body are present near the $\SI{0}{Hz}$, with a very high intensity. Additionally, high-frequency noise originating from the hardware itself is present, with the intensity comparable to the rotor components. Consequently, the rotor micro-Doppler features are not prominent in the spectrogram. 

We apply the proposed rmD-NSP, AM-FM~NSP~\cite{AMFMNSP}, and NSP~\cite{pengNSP} algorithms to extract the UAV rotor signals and perform \ac{set} on the extracted signals. The results are shown in Fig.~\ref{Fig:11}. Fig.~\ref{Fig:11} (a) shows the results after applying the rmD-NSP decomposition. The spectrogram shows that the rotor micro-Doppler signals have high energy, clear features, and distinct periodicity. Within the $\SI{0.1}{s}$ observation period, eight rotations are clearly observed in the spectrogram, which is consistent with the software simulation results shown in Fig.~\ref{Fig:7} (b) and matches the actual rotational speed of $\SI{80}{r/s}$ for the DJI M300 RTK UAV. Additionally, the micro-Doppler frequency caused by the rotor rotation can reach approximately $\SI{4000}{Hz}$, which also aligns with Fig.~\ref{Fig:7} (b). These demonstrate that the ISAC BS can accurately extract UAV's rotor micro-Doppler signals using the proposed algorithm in real urban scenarios. Fig.~\ref{Fig:11} (b) shows the results of extracting rotor signals using AM-FM NSP. The extraction quality is relatively low, capturing only 4-5 rotor rotations within the $\SI{0.1}{s}$ observation period, while retaining some low-frequency vibration components. Fig.~\ref{Fig:11} (c) shows that 
only three rotor rotations are extracted within the $\SI{0.1}{s}$ observation period. Additionally, the strength of the rotor components is comparable to that of the vibration components and high-frequency noise, indicating the poorest extraction performance. The micro-Doppler feature extraction algorithm proposed in this paper can be utilized in complex urban environments. By leveraging Sub-6G base station, it effectively extracts features from rotor signals.

\section{Conclution}

In order to enable the base station to extract micro-Doppler features of aerial targets, this paper proposes a target feature extraction technique. We first considered a more realistic UAV scattering model, taking into account the vibration characteristics, high-speed rotor rotation, and dynamic RCS characteristics. Our goal is to extract the weak UAV's rotor micro-Doppler signals from complex environments and perform feature extraction. In the software simulation, we evaluated the effectiveness of the proposed rmD-NSP algorithm in extracting rotor micro-Doppler signals and compared it with the NSP and AM-FM NSP signal decomposition algorithms. We studied the decomposition performance of this method under different \ac{snr}. Additionally, we validated the algorithm with ISAC hardware testbed results. Using a Sub-$\SI{6}{GHz}$ \ac{isac} hardware testbed, we successfully extracted the rotor micro-Doppler signal and obtained high-resolution features. Within a $\SI{0.1}{s}$ observation period, ISAC BS successfully captures eight rotations of the DJI M300 RTK UAV's rotor in urban environment.

\section{ACKNOWLEDGMENT}
The authors would like to express their sincere gratitude to Dr. Chunwei Meng and Dr. Kan Yu for their invaluable comments and suggestions, which have greatly improved the quality of this paper.

\normalem
\bibliographystyle{IEEEtran}
\bibliography{IEEEabrv,bib/references}

\end{document}